\documentclass[12pt]{article}
\pdfoutput=1

\usepackage[utf8]{inputenc}
\usepackage[left=2.55cm, right=2.55cm, top=2.55cm, bottom=2.55cm]{geometry}
\usepackage{amsmath,amssymb,amsbsy}
\usepackage{slashed}
\usepackage{xcolor}
\usepackage{graphicx}
\usepackage{url}
\usepackage{cancel}
\usepackage{cite}
\usepackage[colorlinks=true,allcolors=blue,pdfborder={0 0 0},linktocpage=false,pdfencoding=auto]{hyperref}
\usepackage{tabularx,booktabs}
\usepackage{multicol}
\usepackage{feynmp}
\usepackage{units}
\usepackage{xspace}
\usepackage[labelfont=bf]{caption}
\usepackage[section]{placeins}
\usepackage{subcaption}
\usepackage{soul} 
\usepackage{bbold}
\usepackage{parskip} 
\usepackage{float}

\definecolor{darkred}{rgb}{0.6,0,0}
\definecolor{darkpurple}{rgb}{0.5,0,0.5}

\newcommand{\code}[1]{\texttt{#1}}

\def\non{\nonumber\\}
\def\B0{B^{(0)}}
\def\A0{A_3^{(0)}}

\def\ea{\epsilon_{1}}
\def\eb{\epsilon_{2}}

\begin{document}

\author{Amin Aboubrahim$^a$\footnote{\href{mailto:aabouibr@uni-muenster.de}{aabouibr@uni-muenster.de}}~, Wan-Zhe Feng$^b$\footnote{\href{mailto:vicf@tju.edu.cn}{vicf@tju.edu.cn}}~, Pran Nath$^c$\footnote{\href{mailto:p.nath@northeastern.edu}{p.nath@northeastern.edu}}~ and Zhu-Yao Wang$^c$\footnote{\href{mailto:wang.zhu@northeastern.edu}{wang.zhu@northeastern.edu}} \\~\\
$^{a}$\textit{\normalsize Institut f\"ur Theoretische Physik, Westf\"alische Wilhelms-Universit\"at M\"unster,} \\
\textit{\normalsize Wilhelm-Klemm-Stra{\ss}e 9, 48149 M\"unster, Germany} \\
$^{b}$\textit{\normalsize Center for Joint Quantum Studies and Department of Physics,}\\
\textit{\normalsize School of Science, Tianjin University, Tianjin 300350, PR. China}\\
$^{c}$\textit{\normalsize Department of Physics, Northeastern University,
Boston, MA 02115-5000, USA} \\}

\title{\vspace{-2cm}\begin{flushright}
{\small MS-TP-21-09}
\end{flushright}
\vspace{2cm}
\Large \bf A multi-temperature universe can allow a sub-MeV dark photon dark matter
\vspace{1cm}}

\date{}
\maketitle

\vspace{1cm}

\begin{abstract}
An   analysis of sub-MeV dark photon as dark matter is given  which is achieved with two
 hidden sectors, one of which interacts directly with the visible sector while the second
 has only indirect coupling with the visible sector. The formalism for the evolution 
 of three  bath temperatures for the visible sector and the two hidden sectors 
 is developed and utilized in solution of Boltzmann equations coupling the three sectors.
 We present exclusion plots where the sub-MeV dark photon can be dark matter.
 The analysis can be extended to a multi-temperature universe with multiple hidden sectors
 and multiple  heat baths.
\end{abstract}

\numberwithin{equation}{section}

\newpage

{ \hypersetup{colorlinks=black,linktocpage=true} \tableofcontents }

\section{Introduction}
\label{sec:intro}

Supergravity and strings models typically contain hidden sectors with gauge
groups including $U(1)$ gauge group factors. These hidden sectors with $U(1)$ gauge
groups can interact feebly with the visible sector and interact feebly or with normal 
strength with each other. The fields in the visible and hidden sectors in general will reside
in different heat baths and the universe in this case will be a multi-temperature
universe.  The multi-temperature nature of the universe becomes a relevant issue if the
observables in the visible sector are functions of the visible and the hidden sector heat 
baths. Such is the situation if dark matter (DM)
resides in the hidden sector but interacts feebly with the visible sector.
 In this case an accurate computation of the relic density requires 
thermal averaging of cross sections and decay widths which depend on temperatures 
of both the visible and the hidden sector heat baths. 
In this work we develop a theoretical formalism which 
can correlate the evolution of  temperatures of the hidden sector and of the visible sectors (for the specific case of two hidden sectors) in an accurate way.

The formalism noted above  is used in 
 the investigation of a dark photon and dark
fermions of  hidden sectors as possible DM candidates.
   There exists a considerable literature in the study of dark photons~\cite{Buckley:2009in,Loeb:2010gj,Kaplinghat:2015aga,Sagunski:2020spe,Aboubrahim:2020lnr,Kaneta:2016wvf,Kaneta:2017wfh,Co:2018lka,Dror:2018pdh,Agrawal:2018vin,Long:2019lwl,AlonsoAlvarez:2019cgw,Nakai:2020cfw,Choi:2020dec,Delaunay:2020vdb,Graham:2015rva,Ema:2019yrd,Ahmed:2020fhc}   
  (for review see~\cite{Tulin:2017ara,Alexander:2016aln,Fabbrichesi:2020wbt}) to which the interested reader is 
  directed. 
 While axions and dark photons in the light to 
 ultralight mass region (from keV to $10^{-22}$ eV)
 have been investigated~\cite{Long:2019lwl,Bloch:2016sjj,Pospelov:2008jk,Nakai:2020cfw,Kim:2015yna,Hui:2016ltb,Halverson:2017deq}, 
  the sub-MeV dark photon mass range appears difficult to realize. The problem
  arises in part because with the visible sector interacting with a hidden sector 
  via kinetic mixing, the twin constraints that the dark photon has a lifetime larger 
  than the age of the universe, and also produce a sufficient amount of DM 
  to populate the universe are difficult to satisfy.
In addition to the relic density constraint there is also the constraint on dark photon  
  lifetime which needs to be larger
than the age of the universe as well as a constraint from BBN on the light degrees of freedom.
For the case of one hidden sector, these constraints are difficult to satisfy.  
Specifically,   $\Delta N_{\rm eff}$ at BBN time  depends on the ratio 
$(T_{\text{hid}}/T_\gamma)^4$. The BBN temperature is typically $\sim  0.1$ MeV and, as will be seen later, for the
case of one hidden sector the ratio $(T_{\text{hid}}/T_\gamma)^4\sim 1$ which gives a contribution to $\Delta N_{\rm eff}$ at BBN 
time in excess of the current experimental constraint which from the combined data from BBN, BAO and 
CMB~\cite{Aghanim:2018eyx} is $\Delta N_{\rm eff} <0.214$.
On the other hand, for the case of two hidden sectors it is possible to satisfy all the current experimental constraints
and for that reason we will focus on the two hidden sector model which is the minimal extension of one hidden
sector model as discussed below.

\begin{figure}[tbp]
\centering
\includegraphics[width=0.75\textwidth]{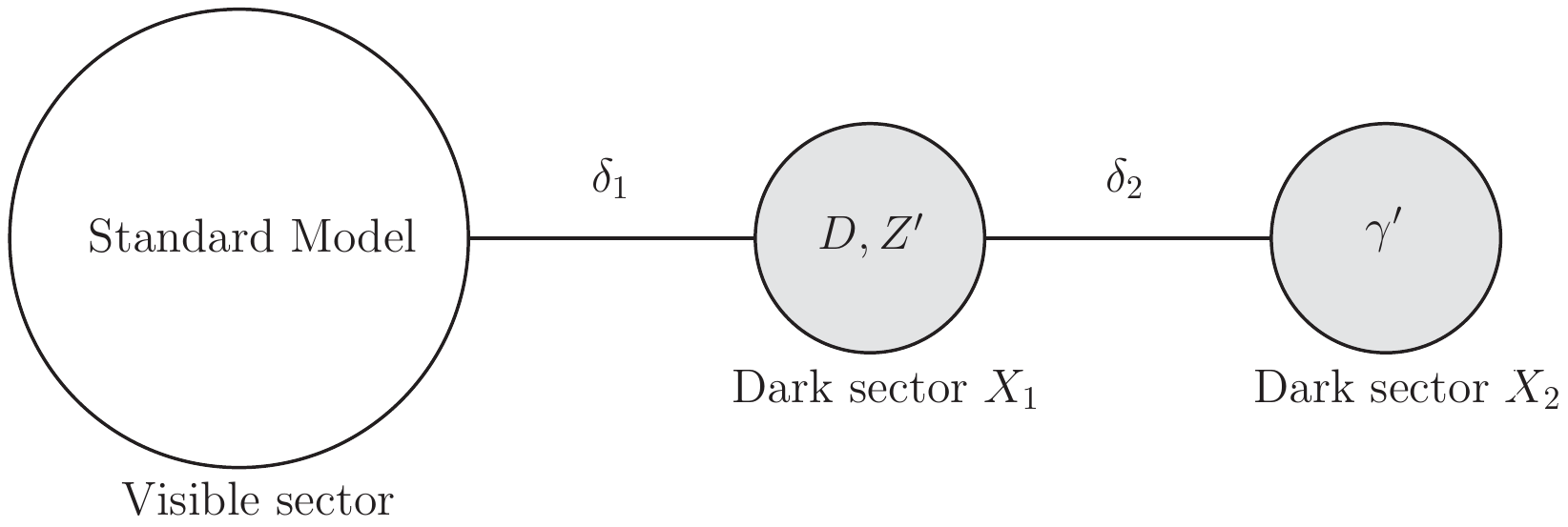}
\caption{The exhibition of the model we consider in the paper. The standard model has a direct coupling with hidden sector $X_1$ with strength proportional to $\delta_1$, whereas hidden sector $X_2$ only interacts directly with $X_1$ with strength proportional to $\delta_2$, and thus $X_2$ interacts with the standard model only indirectly.}
\label{Fig-SM2X}
\end{figure}

 In this study we show that a sub-MeV dark photon as DM can indeed be
 realized in a simple extension of the Standard Model (SM)
 where the hidden sector is constituted of two sectors
 $X_1$ and $X_2$
 where the sector $X_1$ has kinetic mixing with the visible sector while the sector
 $X_2$  has kinetic and mass mixings only
  with sector $X_1$, c.f., Fig.~\ref{Fig-SM2X}.
 We assume that the hidden sector $X_1$ has a dark fermion $D$ and its gauge boson $Z^\prime$ decays
 before the BBN while the hidden sector $X_2$ has only a dark photon $\gamma^\prime$
which has a lifetime larger than the age of the Universe. This set up is theoretically
 more complex because here one has three heat baths and the computation of the
 relic density thus depends on three temperatures, i.e., the temperature $T$ of the visible sector,
 the temperature $T_1$ of the hidden sector $X_1$ and the temperature $T_2$ of the hidden sector
 $X_2$. In the following we develop a formalism that allows one to compute temperatures
 of all three heat baths in terms of one common temperature, which can be chosen to be
 $T$, $T_1$ or $T_2$. In the analysis below it is found convenient to choose 
 the reference temperature to be $T_1$. The outline of the rest of the paper is as follows: in section~\ref{sec:model} we discuss the particle physics model used for our  multi-temperature universe and in sections~\ref{sec:boltzmann} and~\ref{sec:tempevolution} we write the coupled Boltzmann equations with three temperatures and derive the temperature evolution of $T$ and $T_2$ relative to $T_1$. Thermalization between the different sectors and dark freeze-out are explained in section~\ref{sec:numerical} with a numerical analysis and a discussion of the astrophysical constraints. Conclusions are given in section~\ref{sec:conc}. Further analytical results are given in Appendices~\ref{sec:appA} through~\ref{sec:AppD}.

\section{A model for a multi-temperature universe}
\label{sec:model}

\begin{figure}[tbp]
 \centering
   \includegraphics[width=0.75\textwidth]{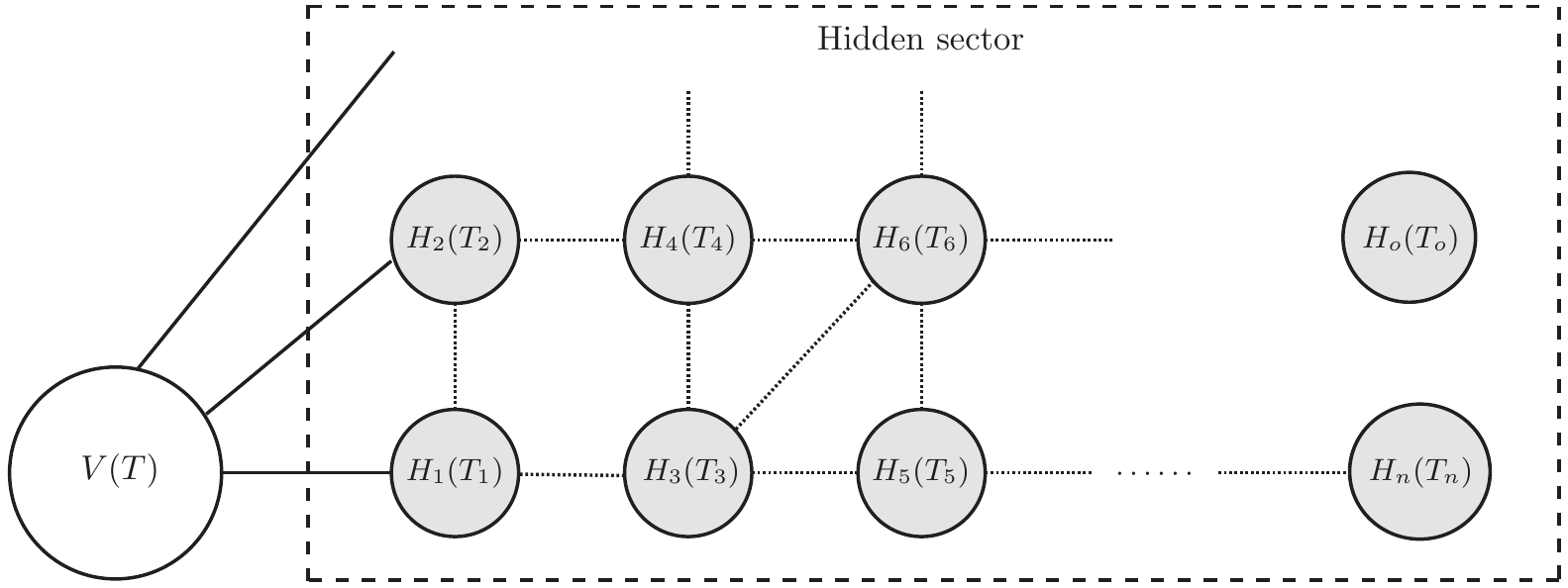}
   \caption{A schematic diagram exhibiting the coupling of the visible sector with multiple dark sectors and of the dark sectors among themselves. 
   The visible sector may have direct couplings with some of the dark sectors, 
   or indirect couplings with others via interactions among the entire hidden sector.}
	\label{Fig-vis-hid}
\end{figure}

As mentioned in the introduction, supergravity and string models
contain hidden sectors.  These hidden sectors in general 
would have both abelian and non-abelian gauge groups
and some of them could interact feebly with the visible 
sector while others may interact with each other
as shown in Fig.~\ref{Fig-vis-hid}. Thus, for example, in D-brane
models one gets $U(N)$ gauge groups where $U(N)\to SU(N)\times U(1)$. These extra $U(1)$ factors in general can acquire kinetic
mixing with the $U(1)_Y$ of the visible sector. Further, the
gauge bosons of the extra $U(1)$'s can acquire mass via the 
Stueckelberg mechanism and also have Stueckelberg mass
mixings with the hypercharge gauge boson.
Additionally if there are several hidden sector $U(1)$'s they 
can have also gauge and Stueckelberg mass mixings among
themselves. Interestingly, the possible existence of these hidden
sectors can have significant effect on model building in the 
visible sector.  As an example one phenomenon which is 
deeply affected by the existence of hidden sectors is dark 
matter which we discuss in further detail below. 

  In general the dark sectors with a gauge symmetry will contain
  gauge fields as well as matter, but, as noted, typically they will have
  feeble interactions with the SM particles and likely also with the 
  inflaton. This means that these particles would not be
  thermally  produced in the reheat period after inflation
  but would then acquire their relic
  density via annihilation and decay of the SM particles. 
  Thus in general the temperatures of the visible and the hidden 
  sectors will be different from the visible sector as
  well as from each other.
  This means that their relic densities will be governed by a set of coupled Boltzmann equations which depend on different
  temperatures, i.e., temperature of the visible sector and those
  for the hidden sectors. One of the central items in understanding
  of how to deal with such coupled systems with sectors involving
  different temperatures is to understand fully how the temperatures
  of the hidden sectors grow relative to the visible sector temperature.
  The formalism of how to correlate the hidden and the visible
  sector temperatures was worked out for the case of the 
    visible sector interacting with one hidden sector in~\cite{Aboubrahim:2020lnr}.
   However, a general framework does not exist. Here 
   we discuss the case where there are two hidden sectors $X_1$
   and $X_2$ where the hidden sector $X_1$ interacts with the
   visible sector, while the hidden sector $X_2$ interacts only 
   with the hidden sector $X_1$ as shown in Fig.~\ref{Fig-SM2X}.   
    In this case two functions $\eta^{-1}=\xi= T_1/T$ and $\zeta = T_2/T_1$  enter in the  coupled Boltzmann equations and we derive differential equations for their evolution. The above setup 
    has a direct application in achieving a sub-MeV dark photon
    as dark matter as we show later in this work. However, a consistent
    analysis of the coupled dynamics of the visible sector and
    two hidden sectors is significantly more complex. This
    work develops the necessary machinery to do so and which can be extended to multiple hidden sectors.

We assume that the two sectors $X_1$ and $X_2$ have 
 $U(1)_{X_1}$ and $U(1)_{X_2}$ gauge symmetries and that the field content of $X_1$ is $(C_\mu, D)$ where 
 $C_\mu$ is the gauge field and $D$ is a dark fermion, and the field content of 
 $X_2$ is the gauge field $D_\mu$
    and there is no dark fermion 
 in the sector $X_2$. 
We invoke a kinetic mixing~\cite{Holdom:1985ag,Dutra:2018gmv}
between the hypercharge field $B_\mu$ of SM and $C_\mu$ and a kinetic mixing
between $C_\mu$ and $D_\mu$ as well as a Stueckelberg mass growth~\cite{Kors:2004dx,Cheung:2007ut,Feldman:2006wb,Feldman:2007wj,Aboubrahim:2019qpc} for all
the gauge fields as well as a Stueckelberg mass mixing between the fields 
$C_\mu$ and $D_\mu$. 
The extended part of the Lagrangian  including both the kinetic and mass mixings is
\begin{equation}
\mathcal{L}=\mathcal{L}_{\rm SM}+\mathcal{L}_{\rm kin}+\mathcal{L}_{\rm mass}+g_X J_{X_1}^{\mu}C_{\mu},
\label{Ltot}
\end{equation}
where $\mathcal{L}_{\rm SM}$ contains the SM terms and the kinetic part is given by
\begin{equation}
\mathcal{L}_{\text{kin}}=-\frac{1}{4}C^{\mu\nu}C_{\mu\nu}-\frac{1}{4}D^{\mu\nu}D_{\mu\nu}-\frac{\delta_1}{2}B^{\mu\nu}C_{\mu\nu}-\frac{\delta_2}{2}C^{\mu\nu}D_{\mu\nu},
\label{Lkin}
\end{equation}
and in the unitary gauge the mass Lagrangian is given by
\begin{equation}
\mathcal{L}_{\text{mass}}=-\frac{1}{2}(M_1 C_{\mu}+M_2B_{\mu})^2-\frac{1}{2}(M_3 C_{\mu}+M_4D_{\mu})^2-m_D \bar{D}D. 
\label{Lmass}
\end{equation}
The $D$ fermion is assumed charged under $U(1)_{X_1}$ with interaction 
$g_X\bar D\gamma^\mu D C_\mu$. Canonical normalization of Eqs.~(\ref{Lkin}) and~(\ref{Lmass}) is carried out in Appendix~\ref{sec:appA} which gives the 
mass eigenstates $\gamma', Z', Z, \gamma$.
The neutral current Lagrangian contained in $\mathcal{L}_{\rm SM}$ for the mass eigenstates 
$\gamma', Z', Z, \gamma$ 
describing the couplings between the vector bosons $\gamma', Z', Z,\gamma$ with the visible sector fermions  is given by
\begin{align}
\mathcal{L}_{\rm NC}^{\rm v}&=\frac{g_2}{2\cos\theta_w}\bar{f}\gamma^{\mu}[(v_f-a_f\gamma_5)Z_{\mu}+(v'_f-a'_f\gamma_5)Z'_{\mu}+(v''_f-a''_f\gamma_5)A^{\gamma'}_{\mu}]f \non 
&~~~+e\bar{f}\gamma^{\mu}Q_f A^{\gamma}_{\mu}f.
\label{NC-visible}
\end{align}    
Here $\theta_w$ is the weak angle and  $e$ is  defined as
\begin{equation}
\frac{1}{e^2}=\frac{1}{g_2^2}+\frac{1+\ea^2-2\ea\delta_1+\ea^2\eb^2-2\ea^2\eb\delta_2}{g^2_Y},
\end{equation}
where $\epsilon_1=M_2/M_1$ and $\epsilon_2=M_3/M_4$.
As seen from Eqs.~(\ref{Ltot})$-$(\ref{Lmass}), the framework of the model allows for the inclusion of both kinetic and mass mixing between the hidden and visible sectors. However, in the analysis presented in this work  we will
only take kinetic mixing between the visible and the hidden sectors, so that $M_2=0$. This is done because
in this case the $4\times 4$ neutral vector boson mass square matrix factors into a block diagonal form consisting of two $2\times 2$ matrices
as shown in Eq.~(\ref{Mst}). In this case we can carry out a set of  eight $GL(2,\mathbb{R})$ transformations to put the kinetic 
energy of the visible and the hidden sectors in  a canonical form  and at the same time to lowest order the mass matrix 
is also in a canonical form. This allows us to use perturbation theory around stable minima of the standard model
mass matrix and therefore to deduce the couplings between the hidden sectors and the SM particles as given
in Tables~\ref{tab1} and~\ref{tab2}.
 As seen in Appendix~\ref{sec:appA}, the analysis is rather non-trivial and significantly more involved
than for the case of one hidden sector.  With the inclusion of $M_2$, the analysis becomes analytically intractable
and an exhibition of results corresponding to those of Tables~\ref{tab1} and~\ref{tab2} is difficult. However, we note that 
even with $M_2=0$, we still have both kinetic and mass mixing in the hidden sector. Thus $\delta_2$ 
takes account of kinetic mixing and $M_3$ takes account of mass mixing between the hidden sectors 1 and 2. 
So in summary in this model the visible sector has only kinetic mixing with the hidden sector 1 while the hidden sector 1 
has both kinetic and mass mixing with hidden sector 2. It is seen that the mass mixing  from $M_3$ does have
significant effect on the model predictions, e.g., on the mass of the dark photon and on the relic density as seen 
in model point (d) in Table~\ref{tab3}.

With $M_2=0$, the
 neutral current Lagrangian for coupling to the 
hidden sector fermion is given by
\begin{align}
\mathcal{L}^{\rm h}_{\rm NC}&=
(c_{\gamma'}A^{\gamma'}_{\mu}+
c_Z Z_{\mu}+c_{Z'} Z'_{\mu}+c_{\gamma}A^{\gamma}_{\mu})
\bar D\gamma^\mu D,
\label{NC-hidden}
\end{align}
where
\begin{align}
\label{ggp}
c_{\gamma'} \simeq&  g_X\dfrac{m^2_{\gamma'}}{m^2_{Z'}-m^2_{\gamma'}}\delta_2, 
~c_{Z}\simeq g_X\delta_1\sin\theta_w (1+\epsilon_z^2),  \\
\label{gzp}
c_{Z'} \simeq &g_X,~
c_{\gamma}\simeq -g_X\delta_1 \delta_2\left(\dfrac{m_{\gamma'}}{m_{Z'}}\right)^2\sin\beta\cos\theta_w, \\
\label{tanb}
\tan 2\beta&=\frac{2 M_3 M_4}{M^2_4-M^2_1-M^2_3}, 
~\epsilon_{z}= m_{Z'}/m_Z.
\end{align}
The vector and axial-vector couplings with SM fermions appearing in Eq.~(\ref{NC-visible}) 
are given by Eqs.~(\ref{vf})$-$(\ref{afpp}). Those couplings along with the ones in Eq.~(\ref{ggp}) and Eq.~(\ref{gzp}) are calculated after a proper diagonalization and normalization of the kinetic and mass square matrices. The complete analysis is given in Appendix~\ref{sec:appA}.

\section{Boltzmann equations for yields with three bath temperatures}
\label{sec:boltzmann}

The relic densities of the dark photon and of the dark fermion arise in part from a freeze-in
mechanism~\cite{Hall:2009bx,Aboubrahim:2019kpb,Aboubrahim:2020wah,Koren:2019iuv,Du:2020avz}.  In general the visible sector and the dark sectors will have different temperatures~\cite{Feng:2008mu,Chu:2011be,Ackerman:mha,Foot:2014uba,Foot:2016wvj,Hambye:2019dwd} (a similar setup has been considered in Ref.~\cite{Hambye:2019dwd} but with a different particle content, couplings and no explicit multi-temperature evolution. Also the DM candidate was not the dark photon as in our case). As mentioned above, we consider three different temperatures corresponding
to the temperatures of the visible sector $T$ and of the two hidden sectors, $T_1$ for 
$X_1$ and $T_2$ for $X_2$.
Defining the yield $Y=n/s$,  where $n$ is the number density and $s$ is the entropy density, 
and the bath functions $\eta$ and $\zeta$ so that 
$T=\eta T_1$ and $T_2=\zeta T_1$, 
we write the Boltzmann equations for the yields as
\begin{align}
\label{boltz-1}
\frac{dY_D}{dT_1}=&-\frac{d\rho/dT_1}{4H\rho}s\mathcal{J}_D,\\
\label{boltz-2}
\frac{dY_{Z'}}{dT_1}=&-\frac{d\rho/dT_1}{4H\rho}s\mathcal{J}_{Z'},\\
\label{boltz-3}
\frac{dY_{\gamma'}}{dT_1}=&-\frac{d\rho/dT_1}{4H\rho}s\mathcal{J}_{\gamma'},
\end{align} 
where $H$ is the Hubble parameter given by
\begin{equation}
H^2=\frac{8\pi G_N}{3}(\rho_v+\rho_1+\rho_2),
\end{equation}
with $\rho=\rho_v+\rho_1+\rho_2$ and $s$ being the energy and entropy densities given by
\begin{align}
\rho&=\frac{\pi^2}{30}\left(g_{\rm eff}^v T^3+g_{1\rm eff} T_1^3+g_{2\rm eff} T_2^3\right), \\
s&=\frac{2\pi^2}{45}\left(h_{\rm eff}^v T^3+h_{1\rm eff} T_1^3+h_{2\rm eff} T_2^3\right),
\label{rho-s}
\end{align}
and the quantities $\mathcal{J}_D$, $\mathcal{J}_{Z'}$ and $\mathcal{J}_{\gamma'}$ are defined in terms of the collision terms as
\begin{align}
 \mathcal{J}_D= \frac{C_D}{s^2}, ~~\mathcal{J}_{Z'}= \frac{C_{Z'}}{s^2}, ~~ \mathcal{J}_{\gamma'}=
\frac{C_{\gamma'}} {s^2}.
\end{align}
The collision terms $C_D$, $C_{Z'}$ and $C_{\gamma'}$ are given by Eqs.~(\ref{jd})$-$(\ref{jgammaprime}) which allow us to write 
 $\mathcal{J}_D, \mathcal{J}_{Z'}$ and $\mathcal{J}_{\gamma'}$ in terms of the yield as
\begin{align}
\mathcal{J}_D&=\langle\sigma v\rangle_{i\bar{i}\to D\bar{D}}(\eta T_1)Y_i^2(\eta T_1)-\frac{1}{2}\langle\sigma v\rangle_{D\bar{D}\to i\bar{i}}(T_1)Y_D^2+\langle\sigma v\rangle_{Z'Z'\to D\bar{D}}(T_1)Y^2_{Z'} \non
&-\frac{1}{2}\langle\sigma v\rangle_{D\bar{D}\to Z'Z'}(T_1)Y_D^2-\frac{1}{2}\langle\sigma v\rangle_{D\bar{D}\to\gamma'\gamma'}(T_1)Y_D^2+\langle\sigma v\rangle_{\gamma'\gamma'\to D\bar{D}}(\zeta T_1)Y_{\gamma'}^2 \non
&-\frac{1}{2}\langle\sigma v\rangle_{D\bar{D}\to Z'\gamma'}(T_1)Y_D^2+\langle\sigma v\rangle_{Z'\gamma'\to D\bar{D}}(T_1,\zeta T_1)Y_{Z'}Y_{\gamma'},    
\end{align}
\begin{align}
\mathcal{J}_{Z'}&=\langle\sigma v\rangle_{i\bar{i}\to Z'Z'}(\eta T_1)Y_i^2(\eta T_1)-\langle\sigma v\rangle_{Z'Z'\to i\bar{i}}(T_1)Y_{Z'}^2-\langle\sigma v\rangle_{Z'Z'\to D\bar{D}}(T_1)Y^2_{Z'} \non
&+\langle\sigma v\rangle_{i\bar{i}\to Z'}(\eta T_1)Y^2_{i}(\eta T_1)+\frac{1}{2}\langle\sigma v\rangle_{D\bar{D}\to Z'Z'}(T_1)Y_D^2+\frac{1}{2}\langle\sigma v\rangle_{D\bar{D}\to Z'\gamma'}(T_1)Y_D^2 \non
&-\langle\sigma v\rangle_{Z'\gamma'\to D\bar{D}}(T_1,\zeta T_1)Y_{Z'}Y_{\gamma'}-\frac{1}{s}\langle\Gamma_{Z'\to i\bar{i}}\rangle(T_1) Y_{Z'},    
\end{align}
\begin{align}
\mathcal{J}_{\gamma'}&=\frac{1}{2}\langle\sigma v\rangle_{D\bar{D}\to\gamma'\gamma'}(T_1)Y^2_D-\langle\sigma v\rangle_{\gamma'\gamma'\to D\bar{D}}(\zeta T_1)Y^2_{\gamma'}+\langle\sigma v\rangle_{i\bar{i}\to\gamma'}(\eta T_1)Y_i^2(\eta T_1) \non
&+\langle\sigma v\rangle_{i\bar{i}\to\gamma'\gamma'}(\eta T_1)Y_i^2(\eta T_1)-\langle\sigma v\rangle_{\gamma'\gamma'\to i\bar{i}}(\zeta T_1)Y_{\gamma'}^2-\frac{1}{s}\langle\Gamma_{\gamma'\to i\bar{i}}\rangle(\zeta T_1) Y_{\gamma'} \non
&+\frac{1}{2}\langle\sigma v\rangle_{D\bar{D}\to Z'\gamma'}(T_1)Y_D^2-\langle\sigma v\rangle_{Z'\gamma'\to D\bar{D}}(T_1,\zeta T_1)Y_{Z'}Y_{\gamma'}.
\end{align}
The thermally averaged cross sections are calculated using Eq.~(\ref{sigmav}) for a single initial state temperature and with Eq.~(\ref{sigv1234}) in the case of two different initial state temperatures, while the thermal averaging of the width is given by Eq.~(\ref{widthav}). Note that the SM effective energy and entropy degrees of freedom ($g^v_{\rm eff}$ and $h^v_{\rm eff}$) are read from tabulated values~\cite{Gondolo:1990dk,Gelmini:1990je}, while those pertaining to the hidden sectors $(g_{1\rm eff},g_{2\rm eff},h_{1\rm eff},h_{2\rm eff})$ are calculated using Eqs.~(\ref{g1eff}),~(\ref{g2eff}) and~(\ref{hdof}). Since the above Boltzmann equations are dependent on the parameters $\eta$ and $\zeta$, then one must consider the evolution of those parameters with temperature. We discuss the formalism in the next section.

\section{Temperature evolution in the dark sectors versus in the visible sector}
\label{sec:tempevolution}

In this section we derive the evolution equations for temperatures
$T_1$ and $T_2$ in the dark sectors and $T$ in the visible sector, i.e., 
$T/T_1$ and $T_2/T_1$ as a function of $T$. However, for 
numerical integration purposes it is found more convenient to 
use $T_1$ as the reference temperature. Thus we are interested
in deriving the evolution equations for $d\eta/dT_1$ and 
$d\zeta/dT_1$ where we recall that $\eta$ and $\zeta$ are defined so that 
\begin{align}
T\equiv \eta T_1,~
T_2\equiv\zeta T_1.
\end{align}
To this end we look at the
equations for the energy densities in the visible and the hidden sectors. In the analysis, we encounter the quantity $d\rho/dT_1$.
The main difficulty in computing the quantity $d\rho/dT_1$ is that
$\rho$ is constituted of three parts,  $\rho=\rho_v + \rho_1+ \rho_2$
which depends on three temperatures i.e., $\rho_v$ is controlled
by $T$, $\rho_1$ is controlled by $T_1$ and $\rho_2$ is controlled
by $T_2$. Thus we need to express $d\rho_v/dT_1$ in terms of 
$d\rho_v/dT$ and $d\rho_2/dT_1$ in terms of $d\rho_2/dT_2$. 
 Using the definitions of $\eta$ and $\zeta$ we can write 
 \begin{align}
 \frac{d\rho_v}{dT_1}= \left(\eta + T_1 \frac{d\eta}{dT_1}\right) \frac{d\rho_v}{dT},~~~ \text{and}~~~
  \frac{d\rho_2}{dT_1}= \left(\zeta+ T_1 \frac{d\zeta}{dT_1}\right) \frac{d\rho_2}{dT}. 
  \label{basic}
\end{align}
This means that a determination of $d\rho_v/dT_1$ and 
$d\rho_2/dT_1$ requires $d\eta/dT_1$ and $d\zeta/dT_2$.
Next, we derive the evolution equations for these quantities. 

We note that $\rho_v, \rho_1, \rho_2$ satisfy the following 
evolution equations
\begin{equation}
\begin{aligned}
\frac{d\rho_v}{dt}+ 4 \rho_v H = j_v, \\
\frac{d\rho_1}{dt}+ 4 \rho_1 H = j_1, \\
\frac{d\rho_2}{dt}+ 4 \rho_2 H = j_2, 
\end{aligned}    
\end{equation}
where $j_v, j_1, j_2$ are the corresponding sources. Instead of time we will use temperature
so we will need to convert derivatives with respect to time to derivatives with respect to 
temperature. We note now that for any given temperatures $T_i$
 the time derivative of temperature is given by 
\begin{align}
\frac{dT_i}{dt}= - \frac{4H\rho}{\frac{d\rho}{dT_i}}\,.
\label{3-8}
\end{align}
As discussed above, we choose $T_1$ to be the reference temperature and the evolution equation for $\rho_v$ in this case can be written as 
\begin{align}
j_v- 4 \rho_v H&= 
- \frac{4H\rho}{\frac{d\rho}{dT_1}}  \frac{d\rho_v}{dT_1}\,.
\label{jv4}
\end{align}
From Eq.~(\ref{jv4}) we can deduce that
\begin{align}
\frac{d\rho_v}{dT_1}&= \frac{4\rho_v H- j_v}{4H (\rho_1+ \rho_2) + j_v}\left(\frac{d\rho_1}{dT_1}  + \frac{d\rho_2}{dT_1}\right).
\label{couple-v}
\end{align}

In a similar fashion starting with the equation for $d\rho_2/dt$ we
can deduce
  \begin{align}
\frac{d\rho_2}{dT_1}&= \frac{4\rho_2 H- j_2}{4H (\rho_1+ \rho_v) + j_2}\left(\frac{d\rho_1}{dT_1}  + \frac{d\rho_v}{dT_1}\right).
\label{couple-2}
\end{align}
Eqs.~(\ref{couple-v}) and~(\ref{couple-2}) are two coupled equations
involving $d\rho_v/dT_1$ and $d\rho_2/dT_1$ which give the 
solution
\begin{align}
\frac{d\rho_v}{dT_1}&= \frac{ (AB+A)}{ (1-AB)} \frac{d\rho_1}{dT_1},
~~~\text{and}~~~\frac{d\rho_2}{dT_1}= \frac{ (AB+B)}{ (1-AB)} \frac{d\rho_1}{dT_1},
\label{rhov2}
\end{align}
where
\begin{equation}
A=  \frac{4\rho_v H- j_v}{4H (\rho_1+ \rho_2) + j_v},~~~
B=  \frac{4\rho_2 H- j_2}{4H (\rho_v+ \rho_1) + j_2}.
\label{AB}
\end{equation}
Using Eqs.~(\ref{basic}) and (\ref{rhov2}) one can then obtain the
relations 
\begin{align}
\frac{d\eta}{dT_1}&= -  \frac{\eta}{T_1} +  \frac{ (AB+A)}{ (1-AB)} \frac{d\rho_1/dT_1}{ 
 T_1\frac{d\rho_v}{dT}},~~~\text{and}~~~
\frac{d\zeta}{dT_1}= -  \frac{\zeta}{T_1} +  \frac{ (AB+B)}{ (1-AB)} \frac{d\rho_1/dT_1}{ 
 T_1\frac{d\rho_2}{dT_2}},
 \label{etazeta}
\end{align}
where
\begin{align}
\frac{d\rho}{dT_1}&=\left(\frac{4H\rho}{4H\rho_1-j_1}\right)\frac{d\rho_1}{dT_1}, ~~~\frac{d\rho_v}{dT}=\frac{\pi^2}{30}\left(\frac{dg_{\rm eff}^v}{dT}\eta^4T_1^4+4g_{\rm eff}^v\eta^3 T_1^3\right), \\
\frac{d\rho_1}{dT_1}&=\frac{\pi^2}{30}\left(\frac{dg_{1\rm eff}}{dT_1}T_1^4+4g_{1\rm eff}T_1^3\right), ~~~\frac{d\rho_2}{dT_2}=\frac{\pi^2}{30}\left(\frac{dg_{2\rm eff}}{dT_2}\zeta^4 T_1^4+4g_{2\rm eff}\zeta^3 T_1^3\right).
\end{align}
Using the fact that $j_v+j_1+j_2=0$, eliminating
 $j_v$ in favor of $j_1$ and $j_2$ and inserting Eq.~(\ref{AB}) in Eq.~(\ref{etazeta}), one can further simplify Eq.~(\ref{etazeta}) to cast $d\eta/dT_1$ and $d\zeta/dT_1$ in their final form as 
\begin{align} 
\label{boltz-4}
\frac{d\eta}{dT_1}=&-\frac{\eta}{T_1}+\left(\frac{4H\rho_v+j_1+j_2}{4H\rho_1-j_1}\right)\frac{d\rho_1/dT_1}{T_1\frac{d\rho_v}{dT}},\\
\frac{d\zeta}{dT_1}=&-\frac{\zeta}{T_1}+\left(\frac{4H\rho_2-j_2}{4H\rho_1-j_1}\right)\frac{d\rho_1/dT_1}{T_1\frac{d\rho_2}{dT_2}}.
\label{boltz-5}
\end{align} 
The source terms $j_1$ and $j_2$ are given by
\begin{align}
j_1=\sum_i[&2Y_i(T)^2 J(i\bar{i}\to D\bar{D})(T)+2Y_i(T)^2 J(i\bar{i}\to Z'Z')(T)+Y_i(T)^2 J(i\bar{i}\to Z')(T) \nonumber \\
&+2Y_{\gamma'}^2J(\gamma'\gamma'\to D\bar{D})(T_2)-\frac{1}{2}Y_{D}^2J(D\bar{D}\to\gamma'\gamma')(T_1)-\frac{1}{2}Y_{D}^2J(D\bar{D}\to Z'\gamma')(T_1) \nonumber \\
&+Y_{Z'}Y_{\gamma'}J(Z'\gamma'\to D\bar{D})(T_1,T_2)]s^2-Y_{Z'}J(Z'\to i\bar{i})(T_1)s,
\label{j1}
\end{align}
\begin{align}
j_2=\sum_i [&Y_i(T)^2 J(i\bar{i}\to\gamma')(T)+Y_D^2 J(D\bar{D}\to\gamma'\gamma')(T_1)-Y_{\gamma'}^2 J(\gamma'\gamma'\to D\bar{D})(T_2) \nonumber \\
&+2Y_i(T)^2J(i\bar{i}\to\gamma'\gamma')(T)-Y_{\gamma'}^2 J(\gamma'\gamma'\to i\bar{i})(T_2)+\frac{1}{2}Y_{D}^2J(D\bar{D}\to Z'\gamma')(T_1) \nonumber \\
&-Y_{Z'}Y_{\gamma'}J(Z'\gamma'\to D\bar{D})(T_1,T_2)]s^2-Y_{\gamma'}J(\gamma'\to \nu\bar{\nu})(T_2)s,
\label{j2}
\end{align}
with
\begin{align}
n_i(T)^2 J(i\bar{i}\to D\bar{D})(T)&=\frac{T}{32\pi^4}\int_{4m^2_D}^{\infty}ds~\sigma_{D\bar{D}\to i\bar{i}}s(s-s_i)K_2(\sqrt{s}/T), \\
n_i(T)^2 J(i\bar{i}\to Z')(T)&=\frac{T}{32\pi^4}\int_{4m^2_i}^{\infty}ds~\sigma_{i\bar{i}\to \gamma'}s(s-s_i)K_2(\sqrt{s}/T), \\
n_D(T_1)^2J(D\bar{D}\to\gamma'\gamma')(T_1)&=\frac{n_D(T_1)^2}{8m^4_D T_1 K_2^2(m_D/T_1)}\int_{4m^2_D}^{\infty}ds~\sigma_{D\bar{D}\to \gamma'\gamma'}s(s-4m^2_D)K_2(\sqrt{s}/T), \\
n_{Z'}J(Z'\to i\bar{i})(T_1)&=n_{Z'}m_{Z'}\Gamma_{Z'\to i\bar{i}}\,.
\label{n2J}
\end{align}
One should not confuse the variable $s$ in Eqs.~(\ref{j1}) and~(\ref{j2}) with the one in Eq.~(\ref{n2J}). The former is the entropy density while the latter corresponds to the Mandelstam variable.  

\section{Thermalization and dark freeze-out}
\label{sec:numerical}

In the analysis we  make certain that the relic density of the dark relics  is consistent
with the Planck data~\cite{Aghanim:2018eyx},
\begin{equation}
\Omega h^2=0.1198\pm 0.0012,
\end{equation} 
along with a $2\sigma$ corridor from theoretical calculations.
Contribution to the relic density arise from $\gamma'$ and $D$ while $Z'$ decays before 
BBN and is removed from the spectrum. 
  In Table~\ref{tab3} we present four benchmarks which satisfy all the experimental constraints. 
  
\begin{table}[tbp]
\centering
\begin{tabular}{|cccccccccc|}
\hline
Model & $m_D$ & $M_1$ & $M_3$ & $M_4$ & $\delta_1$ & $\delta_2$ & $m_{Z'}$ & $m_{\gamma'}$ & $\Omega h^2$ \\
\hline
(a) & 1.00 & 4.50 & 0.0 & 0.43 & $4.0\times 10^{-10}$ & 0.40 & 4.90 & 0.43 & 0.124 \\
(b) & 0.50 & 4.50 & 0.0 & 0.47 & $6.5\times 10^{-11}$ & 0.40 & 4.90 & 0.47 & 0.103 \\
(c) & 0.05 & 4.50 & 0.0 & 0.45 & $5.6\times 10^{-12}$ & 0.40 & 4.91 & 0.45 & 0.102 \\
(d) & 0.62 & 4.50 & -5.0 & 0.45 & $4.0\times 10^{-10}$ & 0.05 & 6.76 & 0.30 & 0.108 \\
\hline
\end{tabular}
\caption{\label{tab3} 
Benchmarks used in this analysis where
 $g_X=0.95$ and masses are in MeV except $m_D$ which is in GeV.
  In the analysis of this table and in the rest of the numerical analysis we choose $M_2=0$. }
\end{table}

The relic density shown is that of $\gamma'$ while that of $D$ is only $\mathcal{O}(10^{-6})$ or less and thus negligible. 
We note here that in addition to the particle physics interactions generating the dark photon relic density, one could
have in addition gravitational production~\cite{Graham:2015rva,Ema:2019yrd,Ahmed:2020fhc}.
 However, such
a production is highly dependent on the  reheat temperature which is model dependent. From Eq.~(46) 
of Ref.~\cite{Graham:2015rva}, for the case when the dark photon mass is $\sim 1$ MeV, one finds that the gravitational production of dark
photon would be suppressed when the reheat temperature $H_I < 10^{11}$ GeV. So one may think of our model
being valid for this restricted class of inflationary models. 

The dark photon is long-lived with a decay width  to two neutrinos given by
\begin{align}
\Gamma_{\gamma'\to \nu\bar{\nu}}&=\frac{g_2^2\delta_1^2(\delta_2-\sin\beta)^2\epsilon_{\gamma'}^4}{8\pi}m_{\gamma'}\tan^2\theta_w\,,
\end{align}
where $\epsilon_{\gamma'}= m_{\gamma'}/m_Z$, and  the partial decay width of $\gamma'$ to three photons reads~\cite{Pospelov:2008jk,McDermott:2017qcg}
\begin{align}
\Gamma_{\gamma^{\prime}\to 3\gamma}& =\frac{17\alpha^{3}\alpha^{\prime}}{2^{7}3^{6}5^{3}\pi^{3}}\frac{m_{\gamma^{\prime}}^{9}}{m_{e}^{8}}\approx4.70\times10^{-8}\alpha^{3}\alpha^{\prime}\frac{m_{\gamma^{\prime}}^{9}}{m_{e}^{8}},
\end{align} 
where $\alpha=e^{2}/4\pi$, $\alpha^{\prime}=(k e)^{2}/4\pi$ and $k=-\delta_{1}(\delta_{2}-\sin\beta)\cos\theta_w$.  The dark photon's lifetime is  larger than the
age of the universe and this is illustrated by model point (d) which gives 
$\tau_{\gamma'\to\nu\bar\nu}\sim 8.4\times 10^{21}$ yrs and $\tau_{\gamma'\to 3\gamma}\sim 5.3\times 10^{15}$ yrs. 

\begin{figure}[tbp]
\centering
\includegraphics[width=0.49\textwidth]{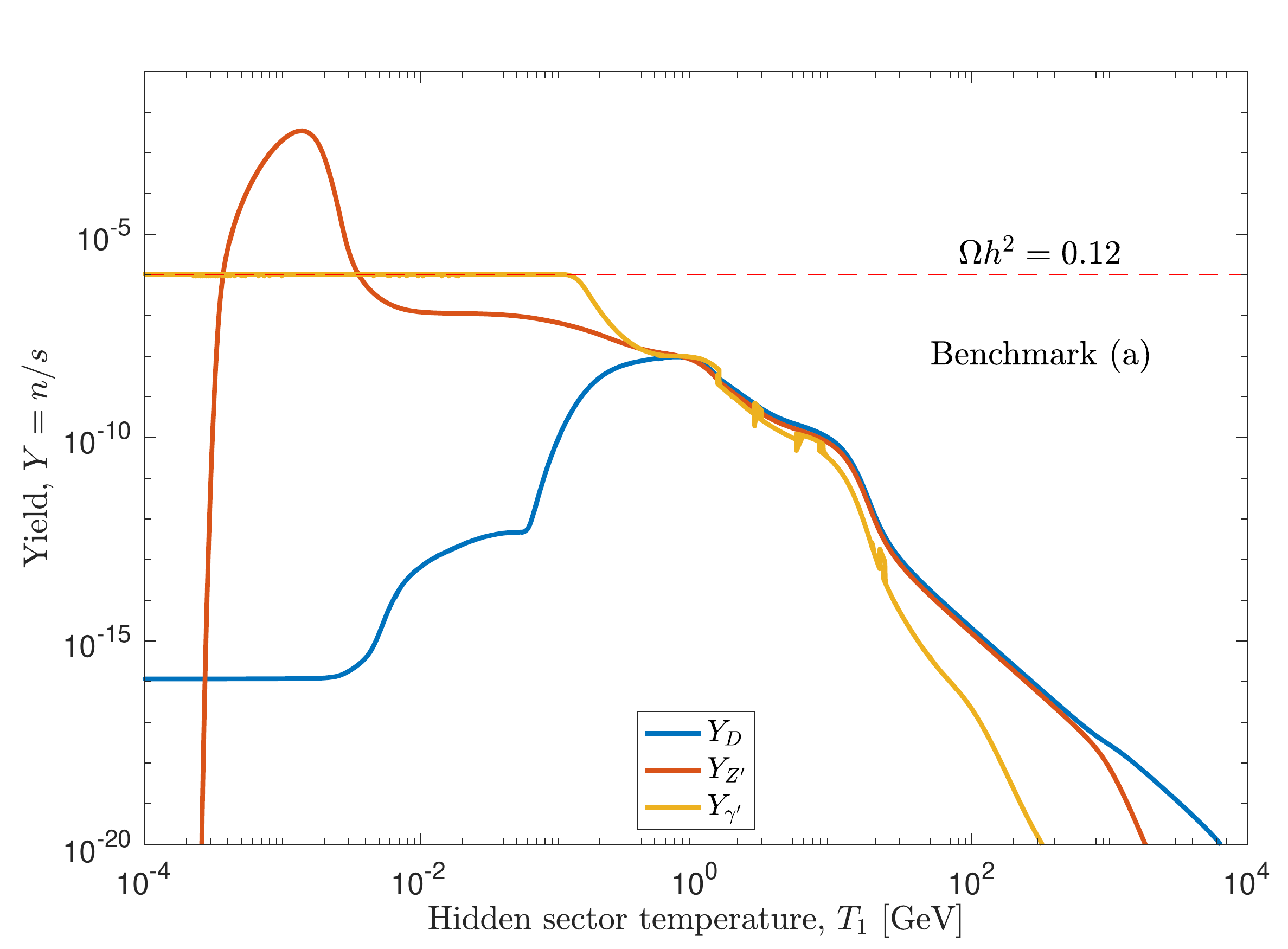}
\includegraphics[width=0.49\textwidth]{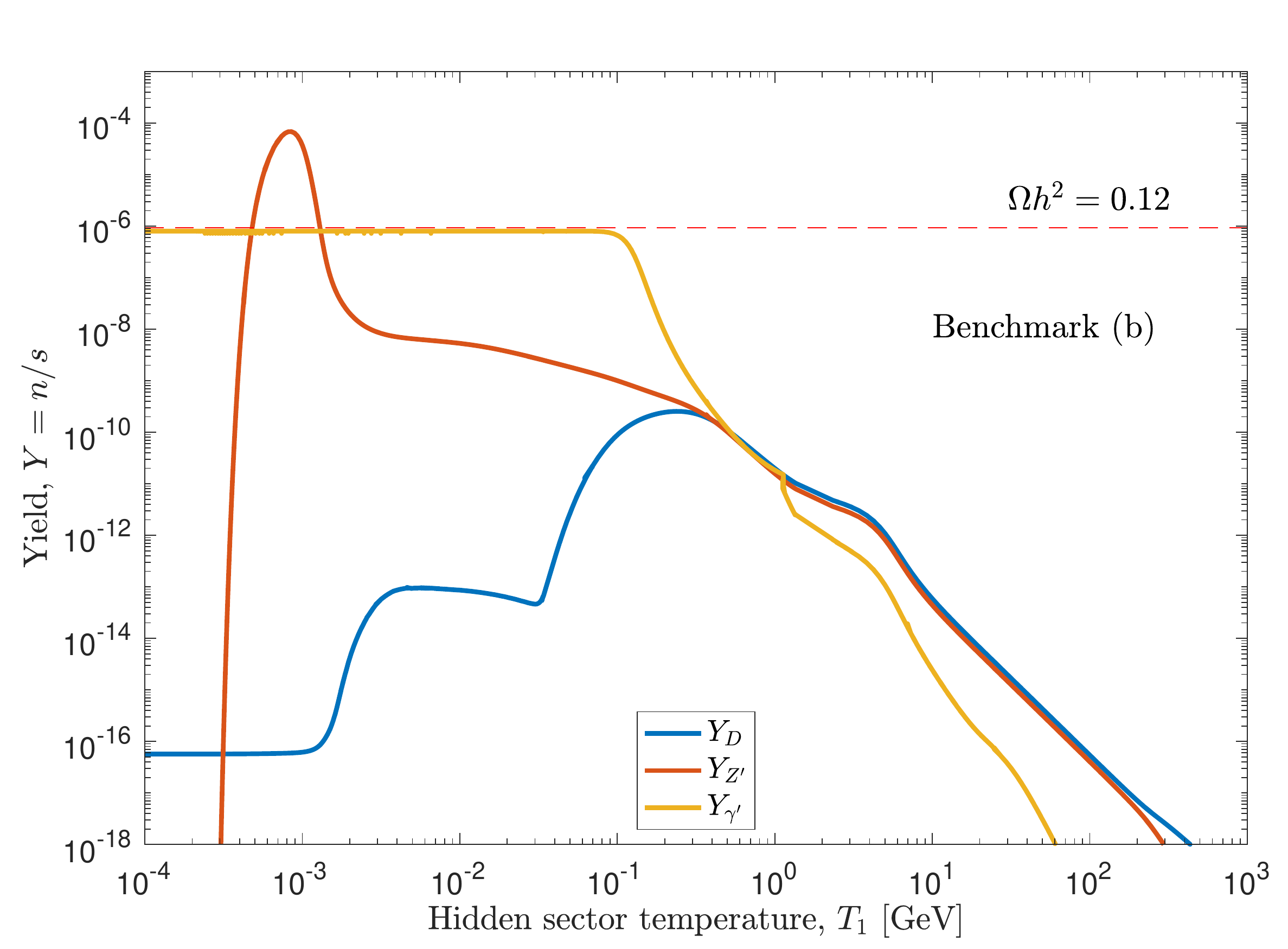}
\caption{The yields for the dark fermion $D$ and the dark bosons $Z'$ and $\gamma'$ as a function of the hidden sector temperature $T_1$ for benchmarks (a) (upper panel) and (b) (bottom panel). The horizontal dashed line corresponds to the observed relic density which matches the freeze-out yield of $\gamma'$. Note that at dark freeze-out $Y_D\ll Y_{\gamma'}$.}
\label{fig1}
\end{figure}

Calculation of the relic density requires determining the yields by numerically solving the five stiff coupled equations, Eqs.~(\ref{boltz-1})$-$(\ref{boltz-3}),~(\ref{boltz-4}) and~(\ref{boltz-5}).
The resulting yields
for  $D$, $Z'$ and $\gamma'$ as a function of the hidden sector temperature $T_1$  are shown in Fig.~\ref{fig1} for benchmarks (a) and (b).
As the universe cools, the number densities of
$D$, $Z^\prime$ and $\gamma^\prime$ increase gradually.
At around $T_1=100$~GeV, the dark fermions $D$ start to freeze-out, 
and the blue curve becomes flat around $T_1=0.1$~GeV.
After the dark fermion decouples, the dynamics of $Z'$ and $\gamma'$ is affected 
mainly by 
SM particles freeze-in processes after $T_1=0.1$~GeV. 
However, since $Z'$ is unstable its density depletes to zero 
at $T_1\sim 10^{-4}$ GeV.  The only particles that contribute to the relic density then
are $D$ (blue curve) and $\gamma'$ (yellow curve).
As noted the analysis gives dark photon as the dominant component of DM. 
We note that the number  changing processes in hidden sector 1 are driven by $D\bar{D}\longleftrightarrow Z'Z'$ owing to the sizable value of the coupling $g_X$. Since $m_{Z'}\ll m_D$, the reaction $Z'Z'\to D\bar{D}$ shuts off early on as the temperature drops while the reverse reaction remains active. This causes a significant drop in $Y_D$ and as a consequence $Y_{Z'}$  rises sharply as shown in Fig.~\ref{fig1}. This is followed by a dramatic drop in $Y_{Z'}$ due to the decay of $Z'$ to SM fermions.

\begin{figure}[H]
\centering
\includegraphics[width=0.49\textwidth]{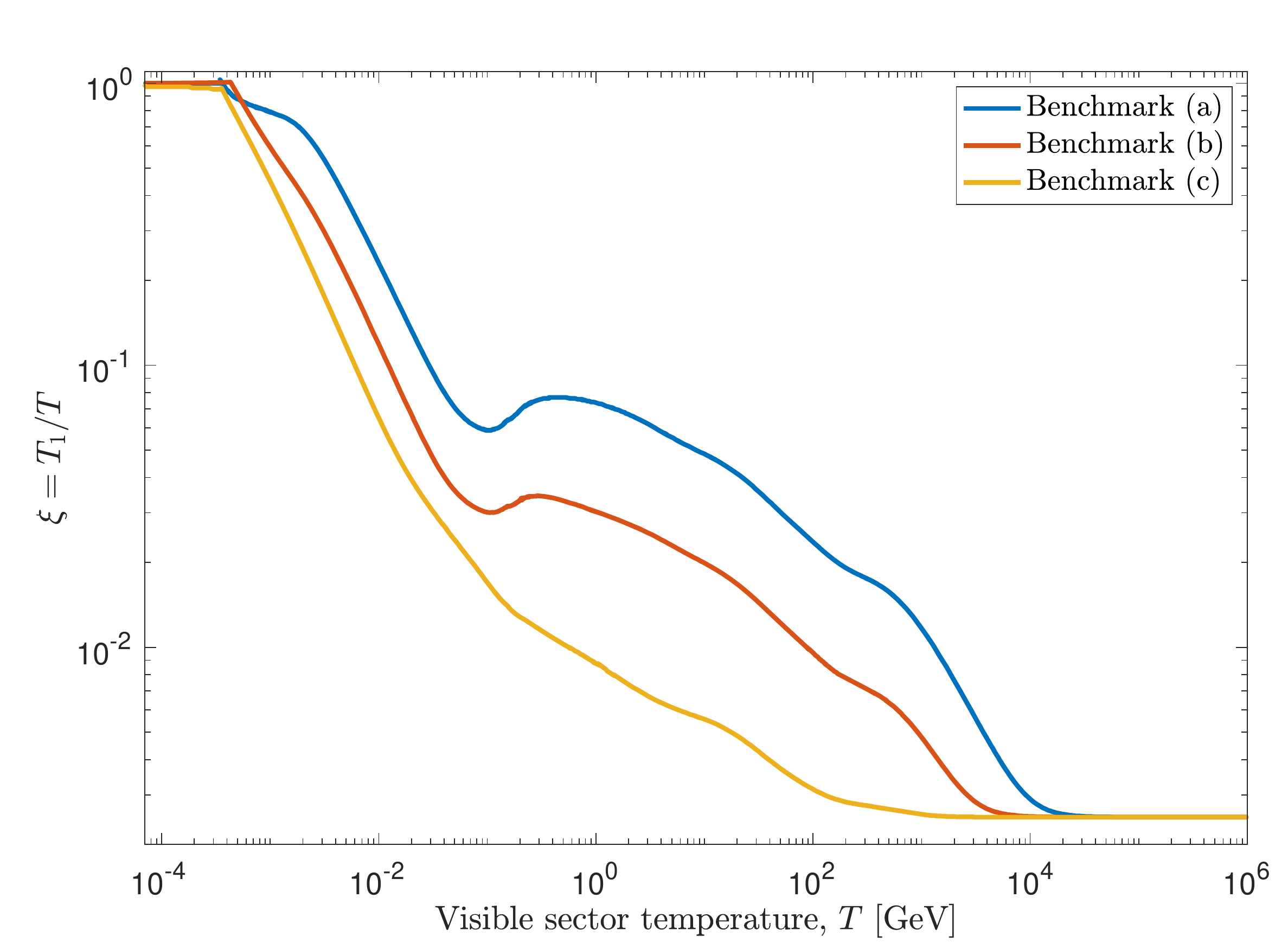}
\includegraphics[width=0.49\textwidth]{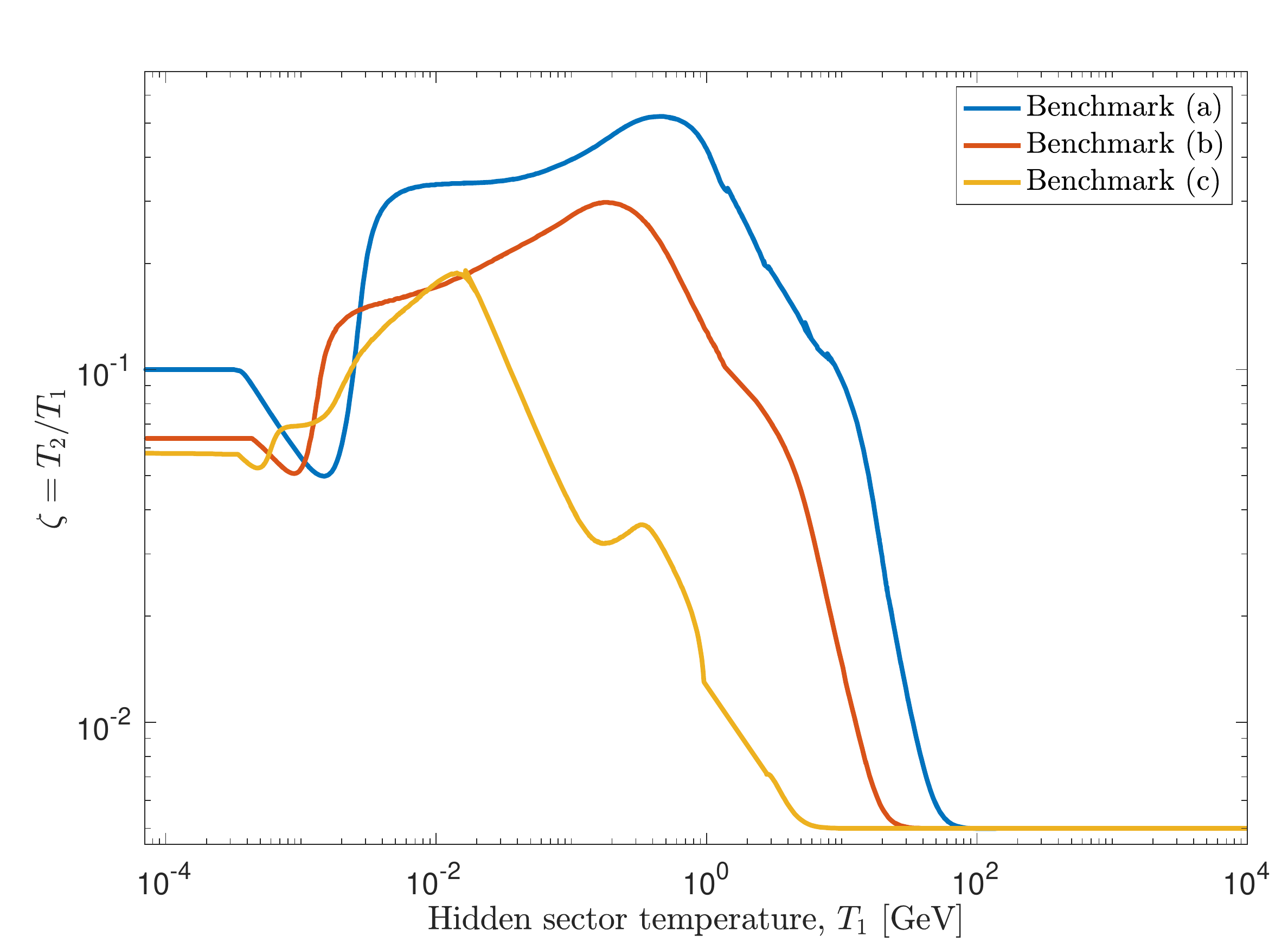}
\includegraphics[width=0.49\textwidth]{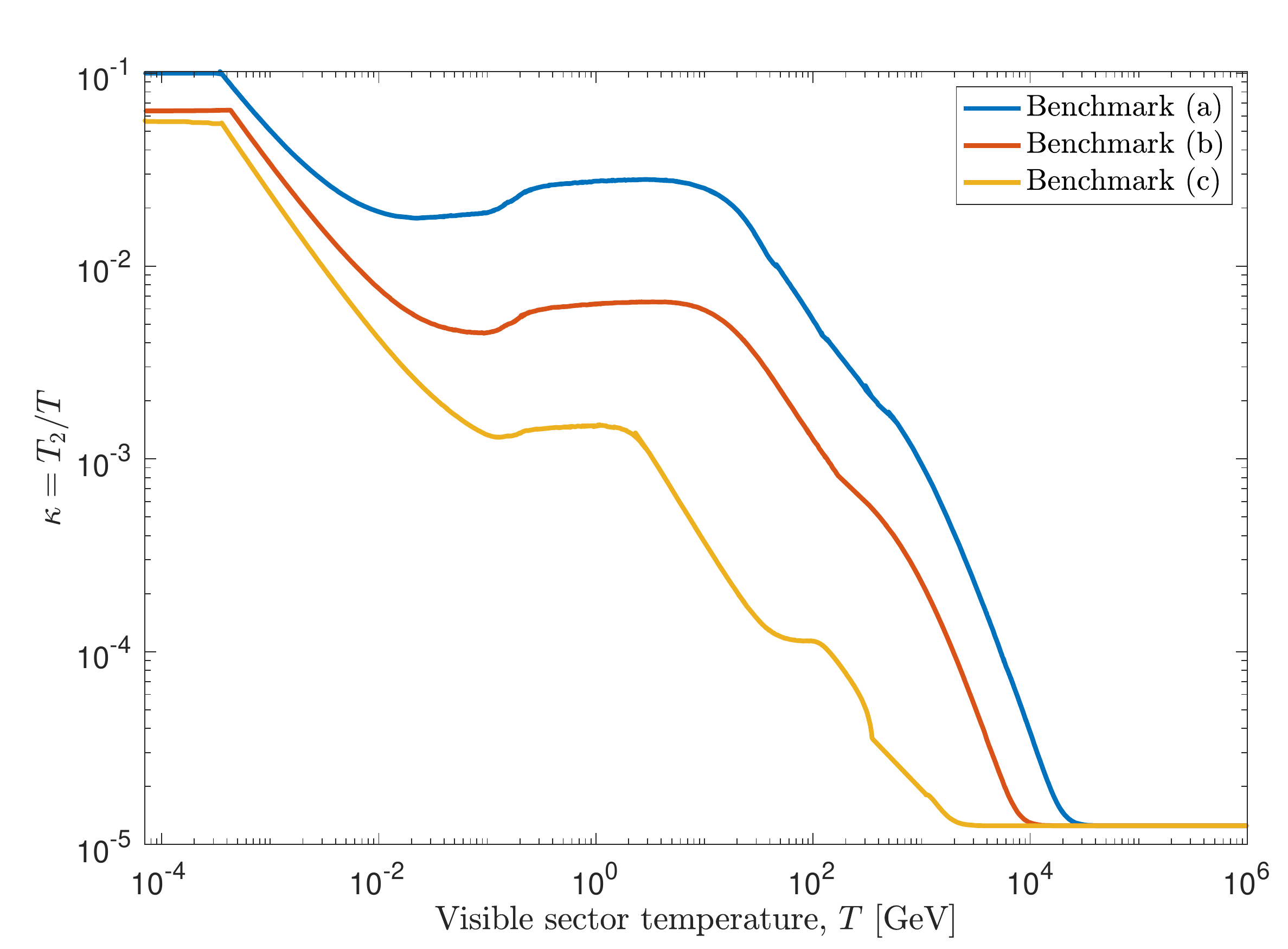}
\caption{Evolution of $\xi$ (upper panel) and $\kappa$ (bottom panel) as a function of the visible sector temperature $T$ and that of $\zeta$ (middle panel) as a function of $T_1$ for three benchmarks (a), (b) and (c) of Table~\ref{tab3}.}
\label{fig2}
\end{figure}

The upper panel of Fig.~\ref{fig2} gives  the evolution of  $\xi =T_1/T$ as a function of 
$T_1$ which shows that $\xi$ rises until it thermalizes with the visible sector, i.e., $\xi\sim 1$.
 The middle panel of this figure gives the evolution $\zeta= T_2/T_1$ as a function of $T_1$ 
  while the bottom panel of Fig.~\ref{fig2} gives
the evolution of $\kappa= T_2/T$ as function of $T$. 
We note that $X_2$ does not thermalize with $X_1$. This happens because  
the energy injection from $X_1$  into $X_2$ is not efficient enough. Consequently 
$T_2/T\ll 1$ which also has  implications for $\Delta N_{\rm eff}$ as we explain later.  

We note in passing that even though hidden sector 1 thermalizes with the visible sector,  there is 
 a distinction between how that happens for the case of the hidden sector versus the visible sector.
In the presence of a coupling induced either by
kinetic mixing or by mass mixing, the visible sector and the hidden sectors will eventually thermalize as long as the 
sectors do not thermally decouple according to the second law of thermodynamics. This is what happens in the top left panel of Fig.~\ref{fig2}. However, we note that the speed with which a dark photon thermalizes is much slower relative to a visible 
sector particle such as a quark which has almost instantaneous thermalization with the photon background.
Further, the thermalization will cease once the particles in the 
hidden sector fully decouple from the visible sector or from each other as seen in the right top panel of Fig.~\ref{fig2}.
This is the case for  hidden sector 2.

\begin{figure}[tbp]
\centering
\includegraphics[width=0.49\textwidth]{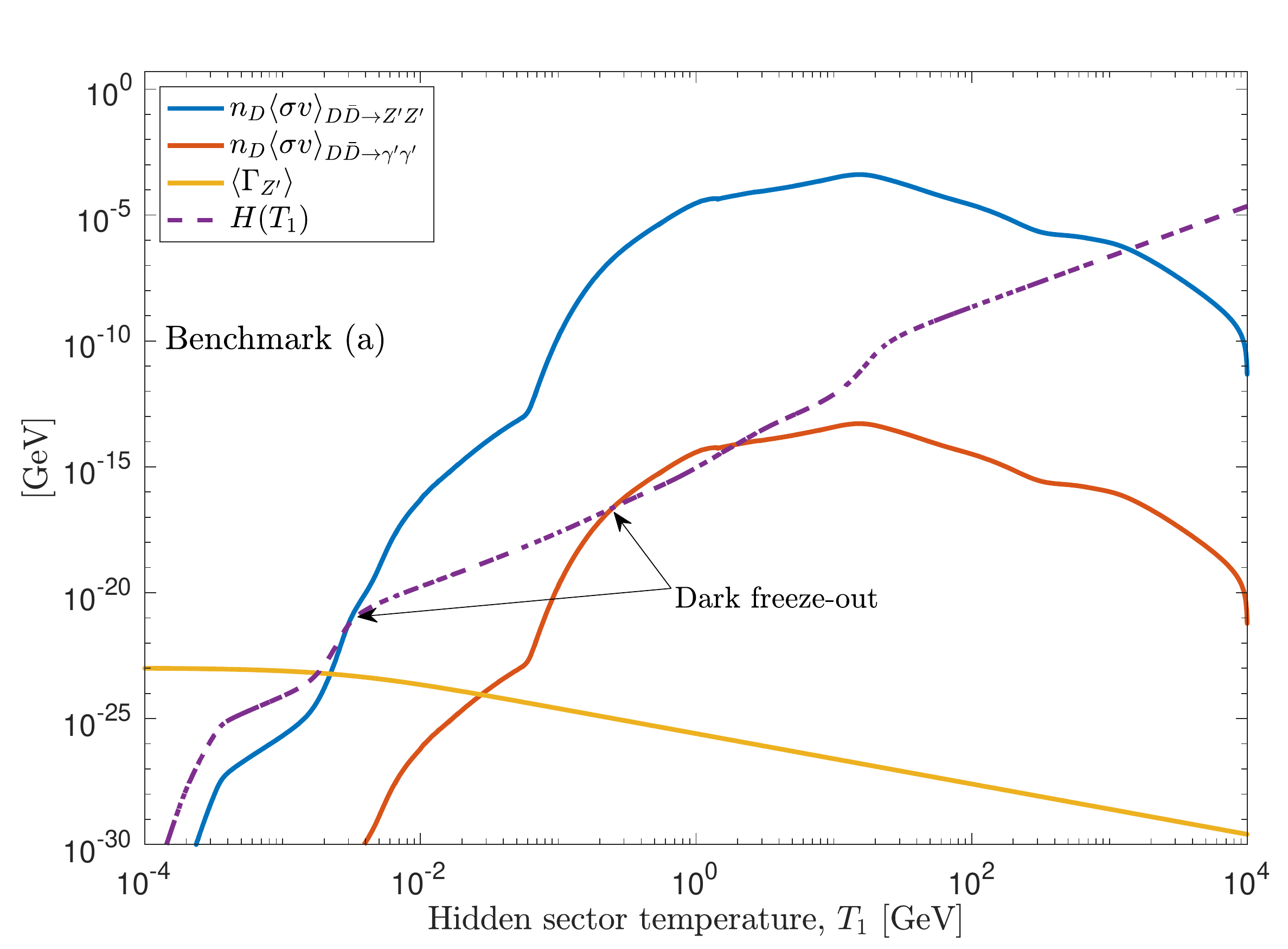}
\includegraphics[width=0.50\textwidth]{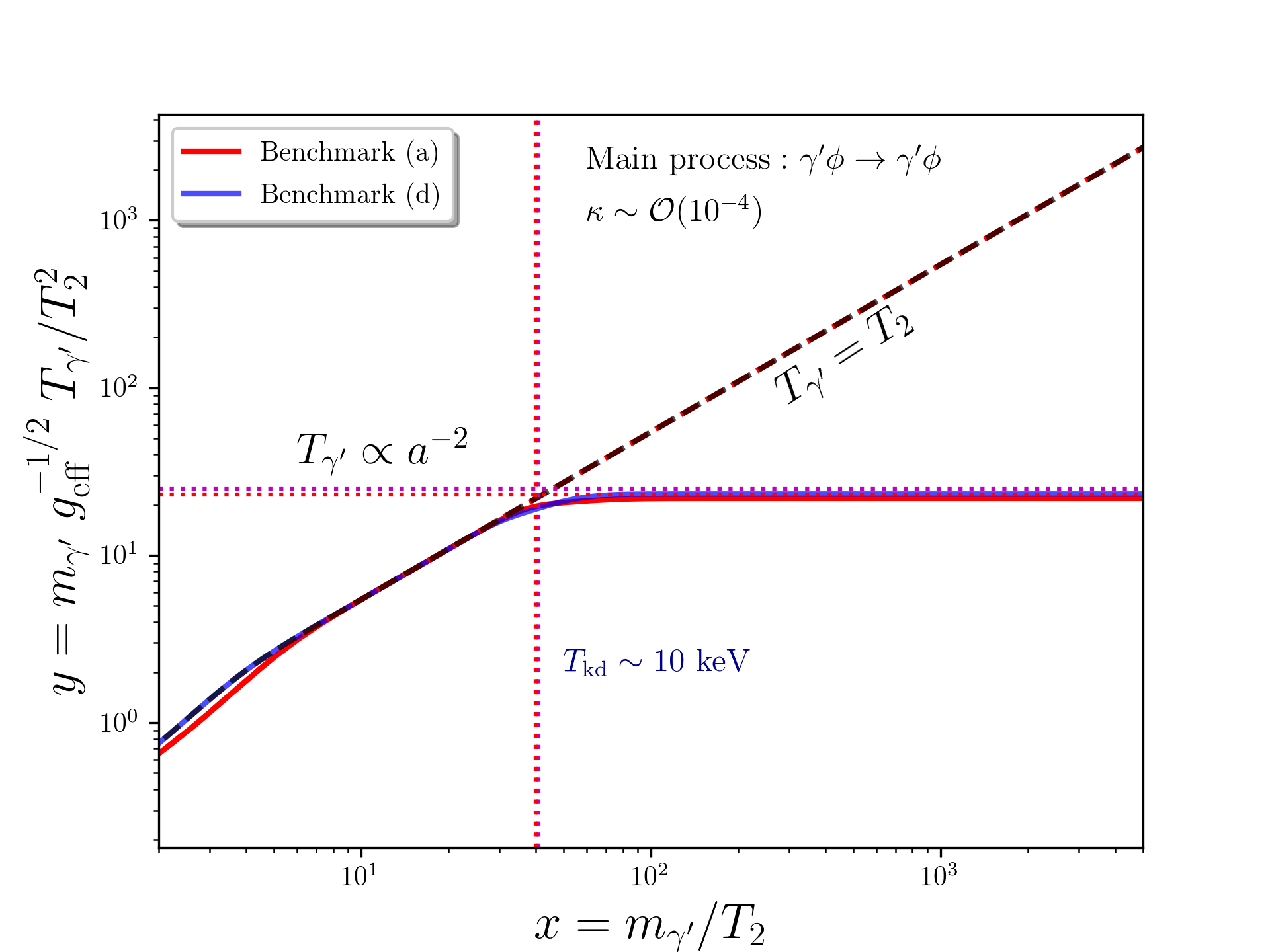}
\caption{Left panel: A plot of $n\langle\sigma v\rangle$ for the dominant processes in the hidden sector along with the Hubble parameter and the thermally averaged decay width of $Z'$. Right panel: the temperature of kinetic decoupling of $\gamma'$ for benchmarks (a) and (d). The dark photon temperature traces that of the thermal bath before decoupling at around 10 keV.}
\label{fig3}
\end{figure}

In the left panel of Fig.~\ref{fig3} we show $n\langle\sigma v\rangle$ and the thermally averaged $Z^\prime$ decay width as a function of $T_1$ for benchmark (a). Also shown is the Hubble parameter $H(T_1)$. As evident, while $Z^\prime$ can enter into equilibrium with the visible sector for a period of time, the dark photon barely does so. We indicate by arrows the point at which the dark freeze-out of $D$ and $\gamma'$ occurs. The dark photons decouple earlier followed by the dark fermions which is also evident in Fig.~\ref{fig1}. We note that $\langle\Gamma_{Z'}\rangle$ overtakes $H(T_1)$ at lower temperatures contributing to the depletion of $Z'$ number density. 
It is of interest to ask how  thermal equilibrium of dark photons occurs once they are produced. Such an equilibrium
can be achieved in our model by considering a massless complex scalar field $\phi$ in the second hidden sector with interactions with the dark photon of the type $(\kappa \phi^{\dagger}\partial^{\mu}\phi A_{\mu}^{\gamma'}+\text{h.c.})$. Elastic scattering between $\gamma'$ and $\phi$, $\gamma'\phi\to\gamma'\phi$, can keep the dark photon in local thermal equilibrium. We follow the method of Ref.~\cite{Bringmann:2009vf}
to determine the temperature of kinetic decoupling of dark photons. For $\kappa\sim 10^{-4}$, we find that kinetic decoupling occurs at 10 keV which is much later than chemical decoupling which happens around 0.1 GeV. 
 The right panel of Fig.~\ref{fig3}  shows the temperature of kinetic decoupling. 
 It is to be noted that the small value of  $\kappa$  has a minimal effect on the relic density of $\gamma'$.

We note that the parameter space of $Z'$ and $\gamma'$ is constrained by experiments such as BaBar, CHARM and other beam-dump experiments as well as by  astrophysical data from Supernova SN1987A and stellar cooling. Those limits become even stronger when the dark photon is assumed to be the dark matter particle. Thus, measurements of heating rates of the Galactic center cold gas clouds~\cite{Bhoonah:2018gjb}, the temperature of the diffuse X-ray background~\cite{Redondo:2008ec} as well as that of the intergalactic medium at the time of He$^{++}$ reionization~\cite{Caputo:2020bdy,Garcia:2020qrp,Witte:2020rvb} are affected by early $\gamma'\to 3\gamma$ decays. Further constraints can be derived from energy injection during the dark ages~\cite{McDermott:2019lch} and spectral distortion of the CMB~\cite{Witte:2020rvb}. The presence of a long-lived sub-MeV particle species can contribute to the relativistic number of degrees of freedom $\Delta N_{\rm eff}$ during BBN and recombination~\cite{Arias:2012az}. All those constraints can exclude a sub-MeV dark photon down to a kinetic mixing coefficient $\mathcal{O}(10^{-13})$. 

\begin{figure}[H]
\centering
\includegraphics[width=0.496\textwidth]{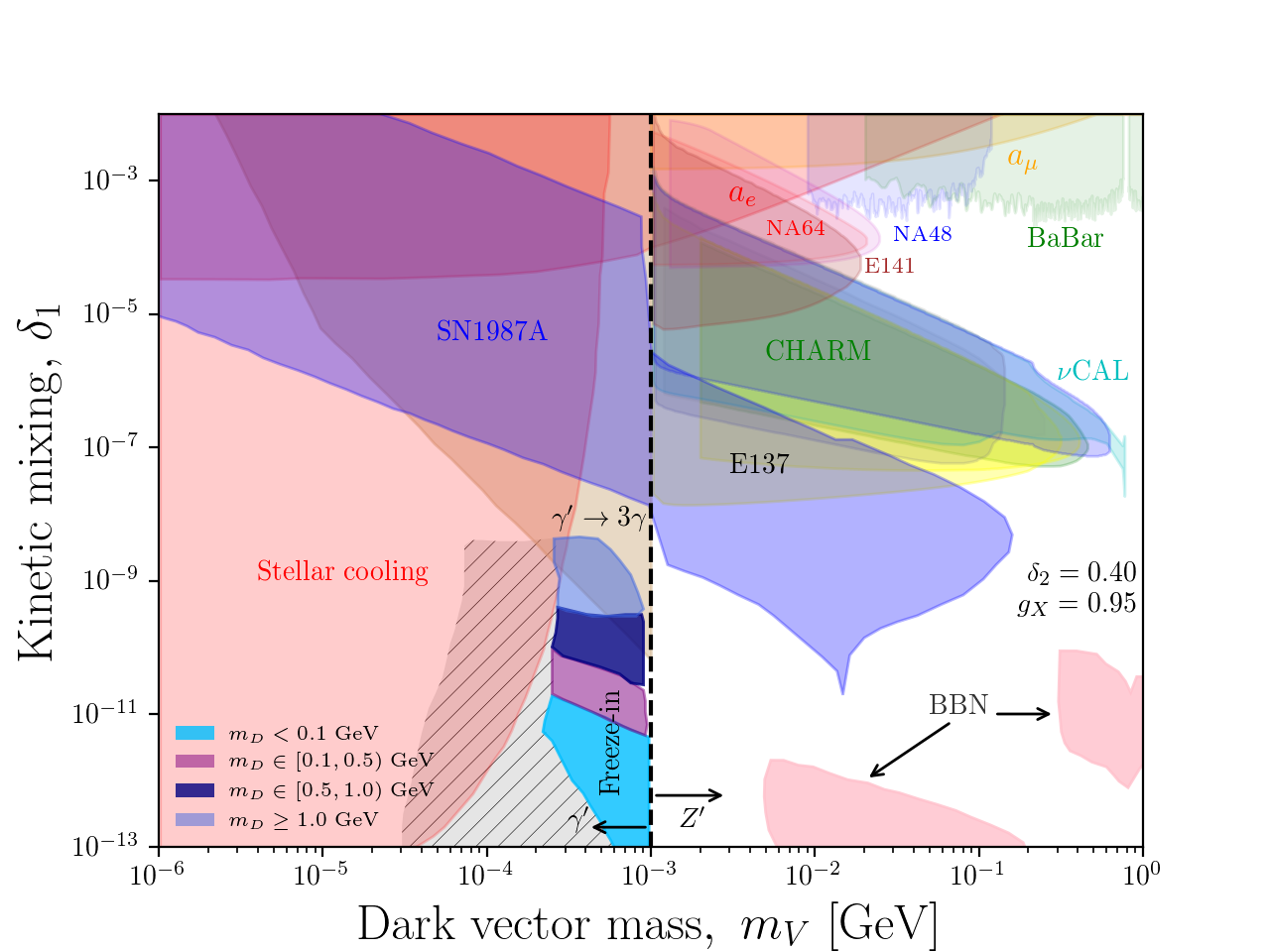}
\includegraphics[width=0.496\textwidth]{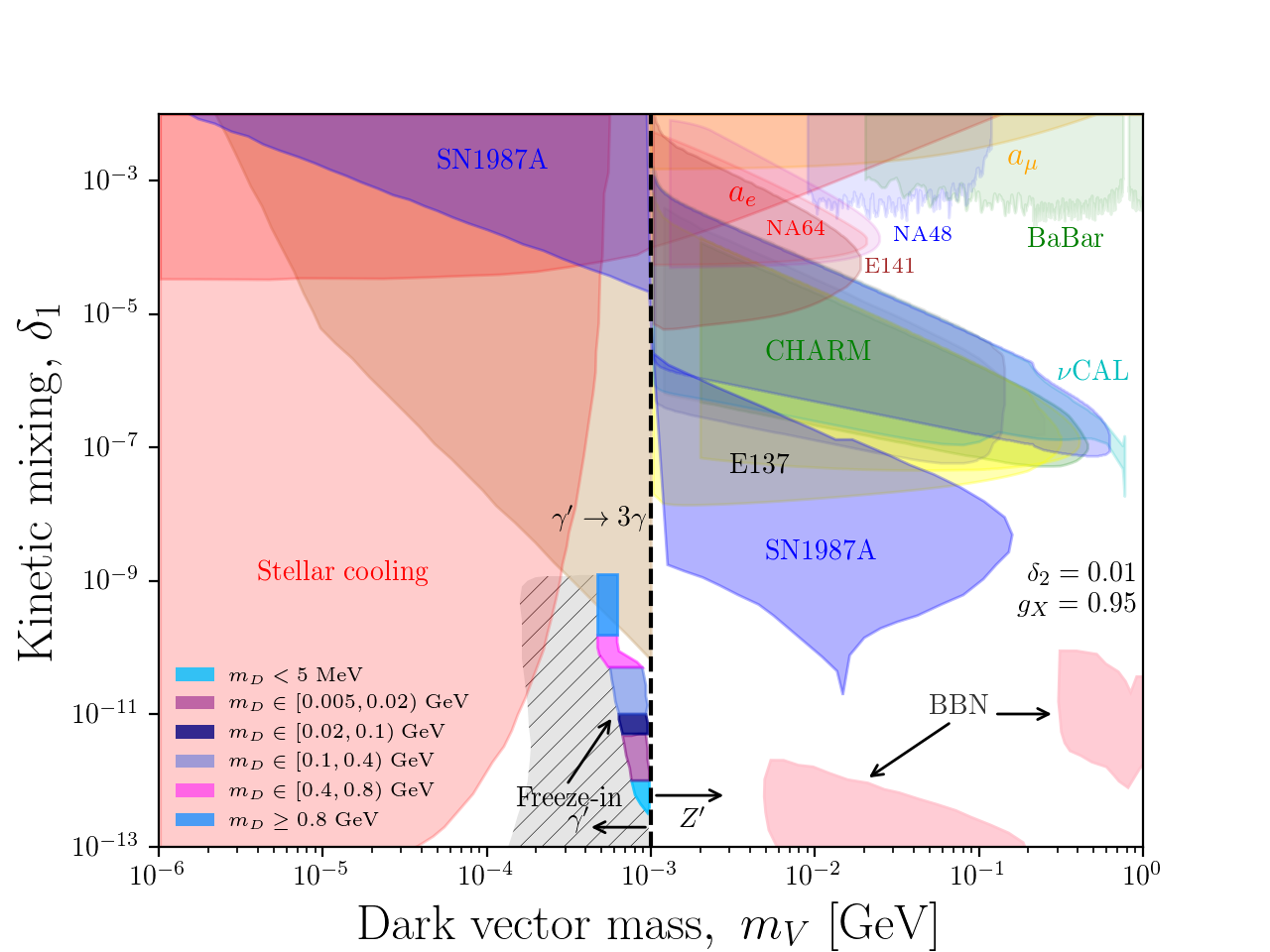}
\caption{Exclusion limits from terrestrial and astrophysical experiments on 
dark photon which has kinetic mixings with the SM sector. Excluded regions are due to
constraints from experiments which include electron and muon $g-2$~\cite{Endo:2012hp}, BaBar~\cite{Lees:2014xha}, CHARM~\cite{Bergsma:1985qz,Tsai:2019mtm}, NA48~\cite{Batley:2015lha}, E137~\cite{Andreas:2012mt,Bjorken:2009mm}, NA64~\cite{Banerjee:2018vgk,Banerjee:2019hmi}, E141~\cite{Riordan:1987aw} and $\nu$-CAL~\cite{Blumlein:1990ay,Blumlein:1991xh,Tsai:2019mtm}. The limits are obtained from \code{darkcast}~\cite{Ilten:2018crw}. The strongest constraints on a dark photon (mass less than 1 MeV) come from Supernova SN1987A (including a robustly excluded region and systematic uncertainties)~\cite{Chang:2016ntp}, stellar cooling~\cite{An:2013yfc} and from decay to $3\gamma$ on cosmological timescales~\cite{Essig:2013goa,Redondo:2008ec}. The islands in pink are constraints from BBN. The region where the freeze-in relic density is satisfied within $2\sigma$ of the experimental constraint is shown in different shades of blue corresponding to different choices of $m_D$. The hatched area represents an enlargement to the $2\sigma$ region allowing a relic density as low as $\sim 10^{-4}$.   In the upper panel, $m_{Z'}=(5-20)m_{\gamma'}$ and $\delta_2=0.4$ while in the lower panel $m_{Z'}=3$ MeV and $\delta_2=0.01$.}
\label{fig4}
\end{figure}

In the model discussed here, the dark photon resides in a hidden sector $X_2$ that does not interact directly with the visible sector. Instead, the direct interaction is between the two hidden sectors $X_1$ and $X_2$ via kinetic and mass mixings. Since $X_1$ mixes kinetically with the visible sector, the interaction between $X_2$ and the visible sector becomes doubly suppressed and all coupling will be proportional to $\delta_1(\delta_2-\sin\beta)$. The quantity $\sin\beta$ is due to the mass mixing between the hidden sector and such a term can impart millicharges to the $D$ fermions. However, the coupling between the photon and the $D$ fermions is not only suppressed by $\delta_1\delta_2\sin\beta$ but also by the mass ratio as evident from the expression of $c_{\gamma}$ 
 in Eq.~(\ref{gzp}). Therefore even for a modest value of $\sin\beta\sim 10^{-2}$, the millicharges are very small and do not constitute a significant constraint on the model.
  This doubly suppressed coupling between the dark photon and the SM can alleviate the present constraints mainly from $\gamma'\to 3\gamma$ as seen in Fig.~\ref{fig4}, which still removes a part of the
parameter space of our model. This constraint is derived from measurements of the intergalactic diffuse photon background. Another decay channel for the dark photon is to two neutrinos. This leads to a possible neutrino flux but experiments have not reached the required sensitivity to probe masses in the sub-MeV region. Experiments such as IceCube~\cite{Aartsen:2018mxl} have constrained only very heavy dark matter decays. The region of the parameter space which would produce a dark photon relic density within $2\sigma$ of the experimental value is shown in both panels of Fig.~\ref{fig4} and labeled `Freeze-in'. For the case when a dominant component of the relic density arises from gravitational production, the parameter space
of our model will be enlarged. The enlarged regions which are represented by the hatched area in Fig.~\ref{fig4}. This area accommodates for a dark photon relic density down to $\sim 10^{-4}$. 

\begin{figure}[tbp]
\centering
\includegraphics[width=0.49\textwidth]{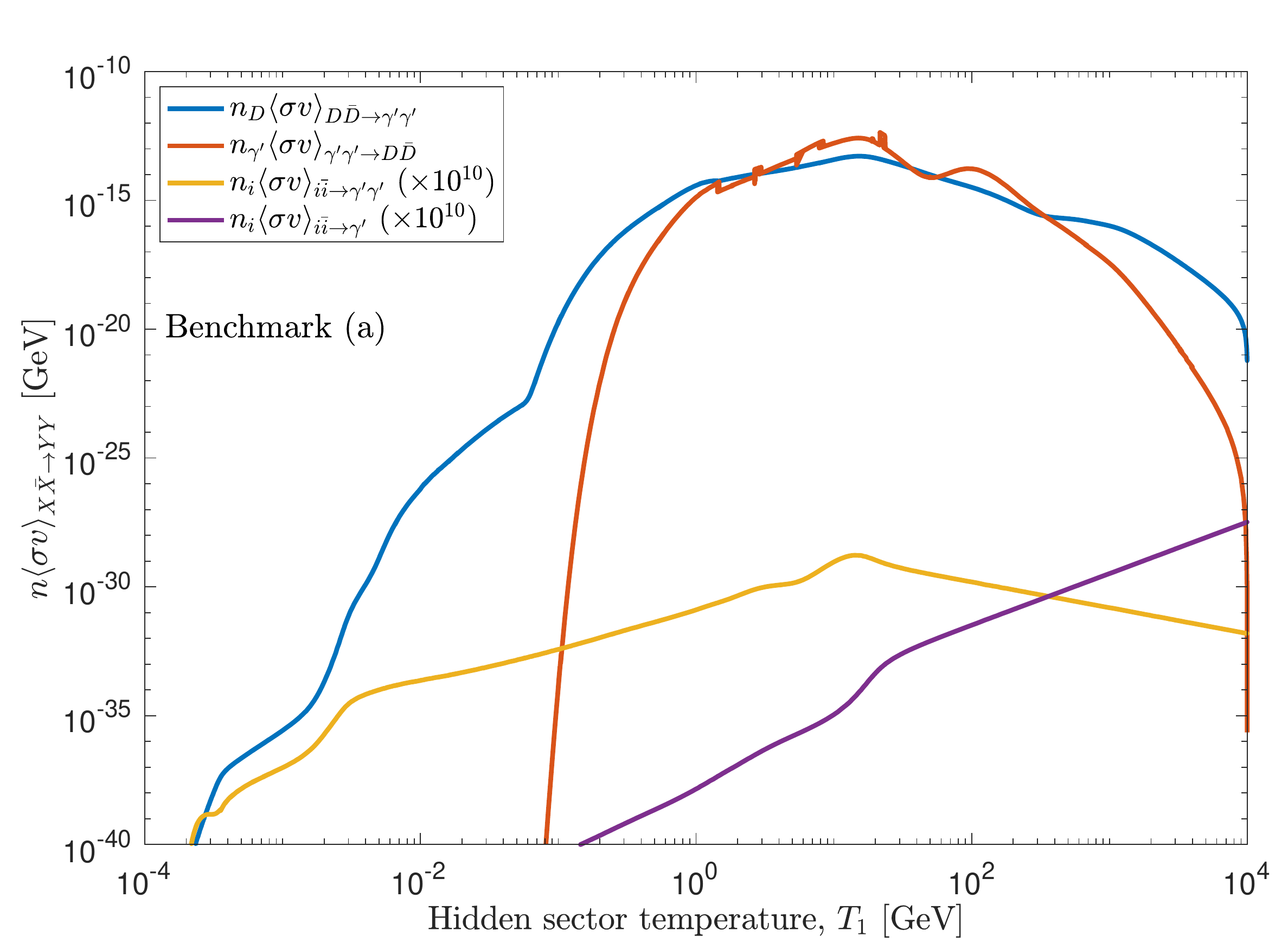}
\includegraphics[width=0.49\textwidth]{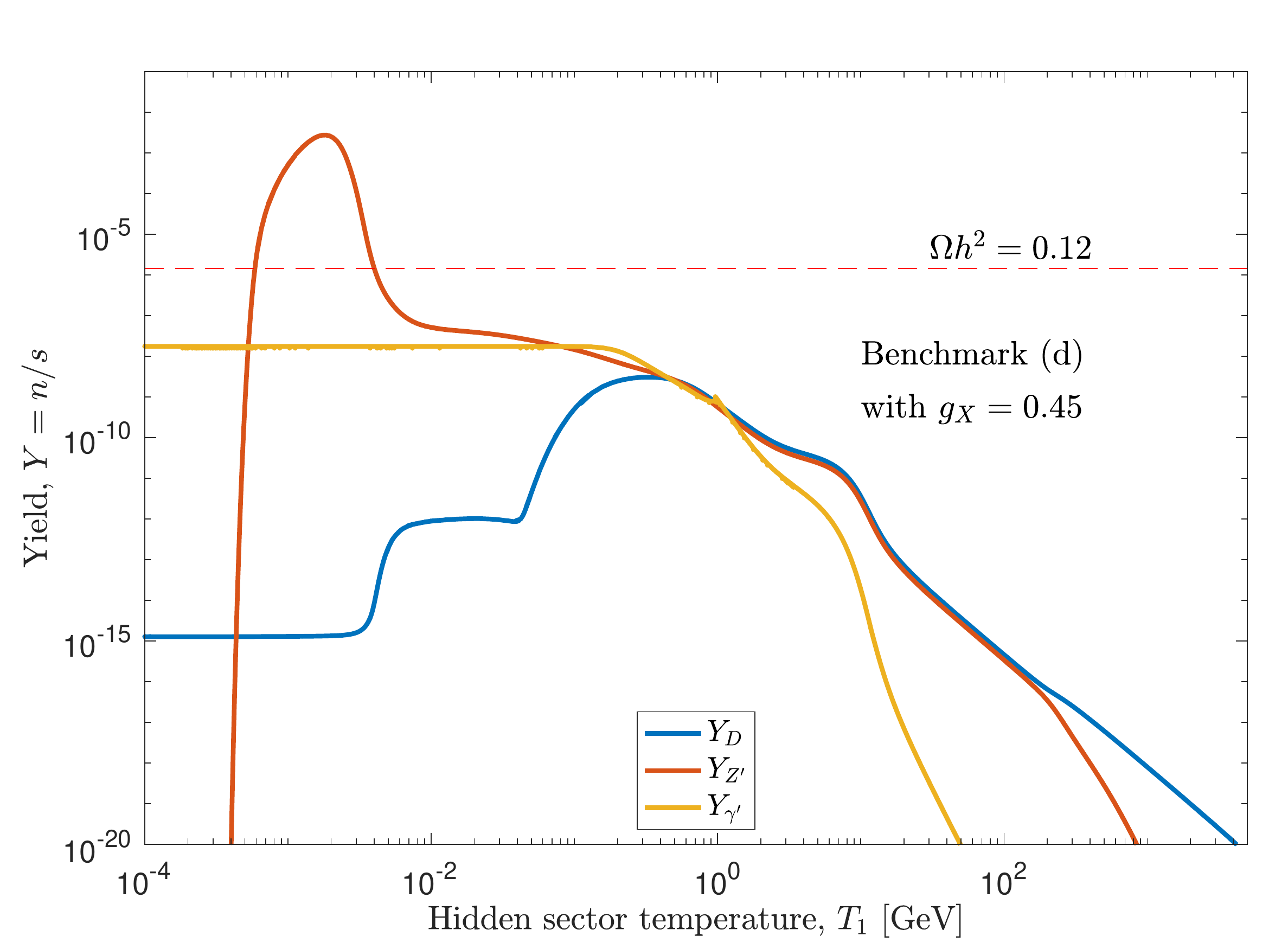}
\caption{Left panel: A plot of $n\langle\sigma v\rangle$ for the processes contributing to the dark photon number density as a function of $T_1$. Right panel: The yields of the hidden sector particles showing a diminishing $Y_{\gamma'}$ due to a smaller $g_X$ for benchmark (d). }
\label{fig1s}
\end{figure}

It is argued in Ref.~\cite{Redondo:2008ec} that a dark photon with direct kinetic mixing with the SM can only give a subdominant contribution to the relic density and that such an observation can be dismissed if another production mechanism is in effect.
The model discussed here presents
exactly this counter argument required to produce a dominant dark photon dark matter. The main production mechanism   
for the dark photon in the current analysis  is not via the freeze-in mechanism from the visible sector, $i\bar{i}\to\gamma'$ and $i\bar{i}\to\gamma'\gamma'$, because of the doubly suppressed coupling (see left panel of Fig.~\ref{fig1s}) but rather from interactions between the hidden sector particles. Thus, processes such as $D\bar{D}\to\gamma'\gamma'$ have cross-sections proportional to $g_X\delta_2$. As shown in Fig.~\ref{fig4}, the sizes of $g_X$ and $\delta_2$ are in the required ranges to produce a dark photon relic density which dominates that of $D$. A smaller value of $g_X$ will reduce the dark photon yield as expected (see right panel of Fig.~\ref{fig1s}). We note in passing that the gauge coupling $g_X$ is constrained by the main annihilation channel $D\bar{D}\to Z'Z'\to 4e$ from the Planck experiment~\cite{Ade:2015xua,Slatyer:2015jla} and for $m_D<1$ GeV, $g_X>0.1$ is excluded. However, this constraint does not exist for our model since the relic abundance of our $D$ fermions is negligible. The effect of the forward process $D\bar{D}\to\gamma'\gamma'$ can be clearly seen in Fig.~\ref{fig1} where the drop in the $D$ fermions yield at a certain temperature is followed by a rise in the yield of $\gamma'$. It is worth mentioning that the reverse process $\gamma'\gamma'\to D\bar{D}$ works on reducing the number density of $\gamma'$ on the expense of $D$, but this process shuts off early on as shown in Fig.~\ref{fig1s} allowing $D\bar{D}\to\gamma'\gamma'$ to completely take over for lower temperatures. 

Finally, we check the number of relativistic degrees of freedom generated by the dark photon (and possibly a complex scalar) at BBN time. The SM gives $N_{\rm eff}=3.046$. The dark photon contribution is given by 
  \begin{align} 
 \Delta N_{\rm eff}\simeq \frac{12}{7}  
 \left(\frac{11}{4}\right)^{\frac{4}{3}} \left(\frac{T_2}{T_{\gamma}}\right)^4,
 \label{Neff}
     \end{align}
where $T_{\gamma}=T$. Using the ratio of the temperatures $T_2/T<0.1$ from Fig.~\ref{fig2}, one finds that  $\Delta N_{\rm eff}$ from dark photons is $\mathcal{O}(10^{-4})$ which makes a negligible contribution
to the SM $N_{\rm eff}$. With the inclusion of a complex scalar, the model has now five new bosonic degrees of freedom but this still gives a small contribution and does not violate the bound on $\Delta N_{\rm eff}$.
We note that the suppressed value of $T_2/T$ which arises from the non-thermalization of 
 sectors 1 and 2 is due to the low density of dark fermions
in sector 1. This can be seen by the yield for the dark fermion in Fig.~\ref{fig1}.

\section{Conclusions}
\label{sec:conc}

  In this work we discussed the possibility that DM in the universe is constituted of 
   sub-MeV dark photons which reside in the hidden sector. In this case 
  a proper analysis of the relic density  requires a
  solution to coupled Boltzmann equations 
  which depend on multiple bath temperatures including the bath temperature for the visible 
  sector and those   for all the hidden sectors that are feebly coupled with the visible sector. 
    In this work we discussed a model
  where the visible sector couples with two hidden sectors $X_1$ and $X_2$ and where
  the particles in the hidden sector consist of a dark fermion, a dark $Z^\prime$ and a dark photon.
  The dark $Z'$ decays and disappears from the spectrum while the dark photon is a long lived 
  relic. 
   We show that the relic density of the dark photon
   depends critically on temperatures of both the visible
   and the hidden sectors. We present exclusion plots where a sub-MeV dark photon can exist
   consistent with all the current experimental constraints. We also show that the 
   existence of a dark photon is consistent with the constraints on $N_{\rm eff}$ from BBN.
   Thus a sub-MeV dark photon is a viable candidate for DM within a constrained
   parameter space of mass and kinetic couplings. 
   The formalism developed here of correlated evolution of bath temperatures in the visible and hidden sectors may find application for a wider class of phenomena involving hidden sectors.

The research of AA was supported by the BMBF under contract 05H18PMCC1. The research of
 WZF was supported in part by the National Natural Science Foundation of China under Grant No. 11905158 and No. 11935009. The research of PN and ZYW was supported in part by the NSF Grant PHY-1913328.

\appendix

\section{Canonical normalization of extended $G_{\text SM}\times U(1)_{X_1}\times U(1)_{X_2}$ Lagrangian with kinetic and Stueckelberg mass mixings}
\label{sec:appA}
 
Consider the Lagrangian 
\begin{equation}
\mathcal{L}=-\frac{1}{4}V_{\mu\nu}^TK_EV^{\mu\nu}-\frac{1}{2}V^TM^2V,
\end{equation}
where $V=(V_1,V_2,V_3,V_4)^T$ which we choose to be $(D,C,B,A_3)$. This Lagrangian can be put in the canonical form
by appropriate transformations on $K_E$ and $M^2$. 
For the case when there is one dark sector it was done 
analytically in~\cite{Feldman:2007wj} and the basic reason which allows that to
happen is that one of the eigenvalues of $M^2$ is zero corresponding
to the photon which effectively reduces the analysis to two massive
modes which can be handled analytically. In the present case 
since we have two hidden sectors,  we have a $4\times 4$ matrix,
and while one of the eigenvalues corresponding to the photon
is zero, one still has to deal with a cubic equation which,
although possible to solve analytically, quickly becomes 
intractable in the presence of both kinetic and Stueckelberg
mass mixing.  However, because the kinetic mixings are typically
small, it is possible to get accurate results by expanding 
couplings of the dark particles with the SM particles in powers
of the kinetic mixings. In this case the relevant couplings can be
recovered easily. However, such an expansion must occur around stable minima. This means that we must first diagonalize
the SM mass squared matrix for the gauge bosons, and compute the
kinetic mixing in this basis. We can then put the kinetic term in
the canonical form. This step requires several  $GL(2,\mathbb{R})$ transformations because of 
several  mixings of the hidden sector with 
the visible sector and the mixing of the two hidden sectors.
 After the kinetic energy is put in the canonical form, we must
 write the mass square matrix of the gauge bosons in the same basis
 which then undiagonalizes the said matrix. However,
because of the smallness of the kinetic mixings, we can carry out
a perturbation expansion of the mass square matrix where
the zeroth order mass square matrix is diagonal and
the perturbations are proportional to the kinetic mixings and are small.
We make this analysis concrete in the formalism below.

Let us consider an orthogonal transformation $V=RV^{(1)}$ such that 
$R^TM^2R=M_D^2$ where $M^2_D$ is a diagonal matrix. In the
$V^{(1)}$ basis the kinetic energy has the form $K_E'= R^T K_E R$.
Next, let us make a transformation $K$ such that $V^{(1)}=K V^{(n)}$,  which could be a product of 
several sub-transformations, such that the kinetic energy is in the canonical form, i.e., 
\begin{equation}
    K^TK_E'K=\mathbb{1}.
\label{matrix-k}
\end{equation}
In our case we will have $n=8$ (as discussed below). 
In the $V^{(8)}$ basis, while the kinetic energy is in the canonical form,
the mass matrix $M^2= K^T M^2_D K$ is not. However, 
as explained above since the kinetic mixings are small we 
can expand $M^2$ around $\delta_1=0=\delta_2$ so that
\begin{align}
K^T M_D^2 K&= M_D^2  + \Delta M^2.
\label{deltams}
\end{align}
Now since the kinetic mixing is supposed to be small, $K$ differs from a unit matrix only
by a small amount and thus $\Delta M^2$ is small relative to $M_D^2$ and one may carry
out perturbation expansion in $\Delta M^2$ to arrive at the kinetic and mass mixing effects
in the physical  processes.
To compute $\Delta M^2$ we need $K$ defined by Eq.~(\ref{matrix-k}).
The computation of $K$ is significantly more complicated than
for the case of one hidden sector. Below we give its computation 
in some detail.

While the procedure outlined above is general we will  discuss the specific
case where  $M^2$ is block diagonal so that
\begin{equation}
M^2=
\begin{pmatrix}
M_4^2&M_3M_4&0&0\\
M_3M_4&M_3^2+M_1^2&0&0\\
0&0&\frac{1}{4}g_Y^2v^2&-\frac{1}{4}g_Yg_2v^2\\
0&0&-\frac{1}{4}g_Yg_2v^2&\frac{1}{4}g_2^2v^2\\
\end{pmatrix},
\label{Mst}
\end{equation}
where the upper right $2\times 2$ matrix is for the hidden sector and the
lower left $2\times 2$ matrix is for the case of the standard model in the
basis $V^T=( D_\mu, C_\mu, B_\mu, A_{3\mu})$. $M^2$ can be diagonalized by $R$
where
\begin{equation}
R=
\begin{pmatrix}
R_{\beta}&0\\
0&R_w\\
\end{pmatrix}, \quad
R_{\beta}=
\begin{pmatrix}
\cos\beta&-\sin\beta\\
\sin\beta&\cos\beta\\
\end{pmatrix}, ~~~~
R_w=
\begin{pmatrix}
\cos\theta_w&-\sin\theta_w\\
\sin\theta_w&\cos\theta_w\\
\end{pmatrix},
\end{equation}
with $\theta_w$ being the weak mixing angle.
Here $R_\beta$ diagonalizes the hidden sector mass squared matrix while
$R_w$ diagonalizes the standard model mass squared matrix, and
the diagonalization gives
\begin{equation}
R^TM^2R=M_D^2\equiv\text{diag}(m_{\gamma'}^2,  m_{Z'}^2, 0,m_{Z}^2).
\end{equation}
Since the mass square matrix is now diagonal, it is a good starting point to diagonalize and normalize the
kinetic energy matrix. This is a bit non-trivial and 
 requires several steps which we outline below. In the basis $(D,C,B,A_3)$, the kinetic Lagrangian is given by
\begin{equation}
\mathcal{L}_{\rm KE}=-\frac{1}{4}(D^2+C^2+B^2+A_3^2)-\frac{1}{4}(2\delta_1BC+2\delta_2CD),
\end{equation}
where we use an abbreviated notation so that 
 $B^2=B_{\mu\nu}B^{\mu\nu}$, $BC=B_{\mu\nu}C^{\mu\nu}$, etc.
Next we write $\mathcal{L}_{\rm KE}$ in the basis $(D^{(1)} , C^{(1)}, B^{(0)}, A_3^{(0)})$ in which the mass square
matrix of the gauge bosons is diagonal. Thus, using
\begin{equation}
 \begin{pmatrix}
D\\
C\\
\end{pmatrix}
=R_\beta\begin{pmatrix}
D^{(1)}\\
C^{(1)}\\
\end{pmatrix},\quad
\begin{pmatrix}
B\\
A_3\\
\end{pmatrix}
=R_w\begin{pmatrix}
\B0\\
\A0\\
\end{pmatrix},
\end{equation}
allows us to write $\mathcal{L}_{\rm KE}$ as
\begin{equation}
\begin{aligned}
\mathcal{L}_{\rm KE}=&-\frac{1}{4}(D^{(1)2}+ C^{(1)2}
+B^{(0)2}+ A_3^{(0)2}) \\
&-\frac{1}{2}\delta_1(B^{(0)}\cos\theta_w-A_3^{(0)}\sin\theta_w)(D^{(1)}\sin\beta+C^{(1)}\cos\beta)\\
&-\frac{1}{2}\delta_2(D^{(1)}\sin\beta+C^{(1)}\cos\beta)(D^{(1)}\cos\beta-C^{(1)}\sin\beta).
\end{aligned}
\label{kin-1}
\end{equation}
The diagonal kinetic terms for 
 $D^{(1)}$ and $C^{(1)}$  have the form 
\begin{equation}
-\frac{1}{4}D^{(1)^2}(1+\delta_2\sin2\beta)-\frac{1}{4}C^{(1)^2}(1-\delta_2\sin2\beta).
\end{equation}
To normalize them to unity we make a transformation 
from the basis  $V^{(1)^T}=  (D^{(1)},C^{(1)},B^{(0)},A_3^{(0)})$ to 
$V^{(2)^T}=  (D^{(2)}, C^{(2)}, B^{(0)}, A_3^{(0)})$
so that
\begin{equation}
   V^{(1)}= K_1 V^{(2)}, ~~  K_1=
\begin{pmatrix}
\frac{1}{\sqrt{1+\delta_2\sin 2\beta}}&0&0&0\\
0&\frac{1}{\sqrt{1-\delta_2\sin 2\beta}}&0&0\\
0&0&1&0\\
0&0&0&1
\end{pmatrix},
\end{equation}
where 
\begin{equation}
\Bar{\delta_2}=\frac{\delta_2\cos2\beta}{\sqrt{1-\delta_2^2\sin^2 2\beta}}.
\end{equation}
After the transformation,   $\mathcal{L}_{\rm KE}$ in the $V^{(2)}$ basis has the form
\begin{equation}
\begin{aligned}
\mathcal{L}_{\rm KE}=&-\frac{1}{4}(D^{(2)^2}+C^{(2)^2}+B^{(0)^2}+A_3^{(0)^2})-\frac{1}{2}\Bar{\delta_2}C^{(2)}D^{(2)} \\
&-\frac{1}{2}(\delta_1^+D^{(2)} +\delta_1^-C^{(2)})(B^{(0)}\cos\theta_w-A_3^{(0)}\sin\theta_w),
\label{c2d2}
\end{aligned}
\end{equation}
where 
\begin{equation}
\delta_1^+=\frac{\delta_1\sin\beta}{\sqrt{1+\delta_2\sin 2\beta}},~~~
\delta_1^-=\frac{\delta_1\cos\beta}{\sqrt{1-\delta_2\sin 2\beta}}.
\end{equation}

We now note that there is a $C^{(2)} D^{(2)}$ mixing term in Eq.~(\ref{c2d2}) which can be removed
by a $GL(2,\mathbb{R})$ transformation. We do this by going from the basis $V^{(2)}$ to 
 $V^{(3)T}=  (D^{(3)},C^{(3)},B^{(0)}, A_3^{(0)})$ so that 
\begin{align}
V^{(2)} =K_2     V^{(3)}, \quad   
    K_2=
\begin{pmatrix}
1&-s_{\Bar{\delta_2}}&0&0\\
0&c_{\Bar{\delta_2}}&0&0\\
0&0&1&0\\
0&0&0&1
\end{pmatrix}, 
\end{align}
where 
\begin{equation}
s_{\Bar{\delta_2}}=\frac{\Bar{\delta_2}}{\sqrt{1-\Bar{\delta_2}^2}},~~~
c_{\Bar{\delta_2}}=\frac{1}{\sqrt{1-\Bar{\delta_2}^2}}.
\label{scdefs}
\end{equation}

In the $V^{(3)}$ basis, $\mathcal{L}_{\rm KE}$
becomes
\begin{equation}
\begin{aligned}
\mathcal{L}_{\rm KE}=-\frac{1}{4}(D^{(3)^2}
+C^{(3)^2}+ B^{(0)^2} +A_3^{(0)^2})&
-\frac{1}{2}\delta_1^-C^{(3)}(B^{(0)} \cos\theta_w-A_3^{(0)}\sin\theta_w)\\&-\frac{1}{2}\delta'D^{(3)}
(B^{(0)} \cos\theta_w-A_3^{(0)}\sin\theta_w),
\end{aligned}
\label{ke3}
\end{equation}
where
\begin{align}
\delta'\equiv\delta_1^+c_{\Bar{\delta_2}}-\delta_1^-s_{\Bar{\delta_2}}.
\end{align}
We note that while there are no kinetic mixing terms between $C^{(3)}$ and $D^{(3)}$, there are kinetic mixing terms
between them and the fields $B^{(0)}$ and $A_3^{(0)}$. The mixing term between
$D^{(3)}$ and $B^{(0)}$ can be removed by the  transformation
\begin{align} 
V^{(3)}=K_3 V^{(4)},~~ K_3=
\begin{pmatrix}
1&0&-s_{\delta_3}&0\\
0&1&0&0\\
0&0&c_{\delta_3}&0\\
0&0&0&1
\end{pmatrix},
\end{align}
where $V^{(4)^T}= (D^{(4)}, C^{(3)}, B^{(1)}, A_3^{(0)})$, $s_{\delta_3}$ and $c_{\delta_3}$ are defined similar to Eq.~(\ref{scdefs}) and $\delta_3$ is defined by 
\begin{equation}
\delta_3\equiv \delta'\cos\theta_w\,.
\end{equation}
 
In the $V^{(4)}$ basis the Lagrangian takes the form 
\begin{equation}
\begin{aligned}
\mathcal{L}_{\rm KE}=&-\frac{1}{4}(D^{(4)^2}
+C^{(3)^2}+ B^{(1)^2} +A_3^{(0)^2})
-\frac{1}{2}\delta_1^-C^{(3)}(B^{(1)} c_{\delta_3}\cos\theta_w-A_3^{(0)}\sin\theta_w) \\
&-\frac{1}{2}\delta'D^{(4)}
(B^{(1)} c_{\delta_2} \cos\theta_w-A_3^{(0)}\sin\theta_w)
+ \frac{1}{2}\delta' s_{\delta_3} B^{(1)}
(-B^{(1)} s_{\delta_3} \cos\theta_w-A_3^{(0)}\sin\theta_w).
\end{aligned}
\label{ke4}
\end{equation}
 
A mixing term between $D^{(4)}$ and $A_3^{(0)}$  exists which can be removed 
  by the transformation
\begin{align} 
V^{(4)}=K_4 V^{(5)},~~ 
    K_4=
\begin{pmatrix}
1&0&0&-s_{\delta_4}\\
0&1&0&0\\
0&0&1&0\\
0&0&0&c_{\delta_4}
\end{pmatrix},
\end{align}
 where $V^{(5)}$ is given by  $V^{(5)^T}= (D^{(5)}, C^{(3)}, B^{(1)}, A_3^{(1)})$ 
and where $s_{\delta_4}$ and $c_{\delta_4}$ are defined as in Eq.~(\ref{scdefs}) and 
\begin{equation}
\delta_4\equiv-\delta'\sin\theta_w\,.
\end{equation}

In the $V^{(5)}$ basis,  $\mathcal{L}_{\rm KE}$ takes the
form
\begin{align}
\mathcal{L}_{\rm KE}=-\frac{1}{4}(D^{(5)^2}+C^{(3)^2}
+B^{(1)^2}+ A_3^{(1)^2})&
-\frac{1}{2}(B^{(1)}c_{\delta_3}\cos\theta_w-A_3^{(1)}c_{\delta_4}\sin\theta_w)\delta_1^-C^{(3)} \non
&-\frac{1}{2}\sin\theta_w\delta'\delta_3 c_{\delta_4} A_3^{(1)} B^{(1)}.
\end{align}

Next we look at the kinetic mixing of  $(C^{(3)}, B^{(1)})$.
This mixing can be removed by the transformation  
\begin{align}
V^{(5)}= K_5 V^{6},  ~~~K_5=
\begin{pmatrix}
1&0&0&0\\
0&1&-s_{\delta_5}&0\\
0&0&c_{\delta_5}&0\\
0&0&0&1
\end{pmatrix}.
\end{align}
Here $V^{(6)^T}= (D^{(5)}, C^{(4)}, B^{(2)}, A_3^{(1)})$ and 
\begin{equation}
\delta_5=\delta_1^-c_{\delta_3}\cos\theta_w\,,
\end{equation}
where $s_{\delta_5}$ and $c_{\delta_5}$ are defined as in Eq.~(\ref{scdefs}).
After the transformation the kinetic energy Lagrangian in the $V^{(5)}$ basis has the form
\begin{equation}
\begin{aligned}
\mathcal{L}_{\rm KE}=&-\frac{1}{4}( D^{(5)^2}+ 
C^{(4)^2}+B^{(2)^2} + A_3^{(1)^2})
+ \frac{1}{2}\sin\theta_w\delta_1^- c_{\delta_4}A_3^{(1)}(C^{(4)}-s_{\delta_5} B^{(2)}) \\
&-\frac{1}{2}\sin\theta_w\delta'\delta_3 c_{\delta_4}c_{\delta_5} A_3^{(1)} B^{(2)}.
\end{aligned}
\end{equation}

Next we examine  the kinetic mixing of the fields $(C^{(4)},A_3^{(1)})$.
This mixing term can be eliminated by the transformation 
\begin{align}
V^{(6)}= K_6 V^{(7)}, ~~~    K_6=
\begin{pmatrix}
1&0&0&0\\
0&1&0&-s_{\delta_6}\\
0&0&1&0\\
0&0&0&c_{\delta_6}
\end{pmatrix},
\end{align}
where  $V^{(7)^T}= (D^{(5)}, C^{(5)}, B^{(2)}, A_3^{(2)})$ and where $\delta_6$ is defined by
\begin{equation} 
\delta_6=-\delta_1^-c_{\delta_4}\sin\theta_w,
\end{equation}
and $s_{\delta_6}$ and $c_{\delta_6}$ are defined as usual. 

After the transformation the kinetic energy Lagrangian in the $V^{(7)}$ basis
has the form
\begin{equation}
\mathcal{L}_{\rm KE}=-\frac{1}{4}( D^{(5)^2}+ 
C^{(5)^2}+B^{(2)^2} + A_3^{(2)^2})
+ \frac{1}{2} 
\delta_7B^{(2)} A_3^{(2)},
\end{equation}
where
\begin{equation}
\delta_7=\sin\theta_w\delta_1^-c_{\delta_4}c_{\delta_6}s_{\delta_5}+\sin\theta_w\delta'c_{\delta_4}c_{\delta_6}c_{\delta_5}s_{\delta_3}.
\end{equation}

We are now left with the last kinetic mixing term involving the fields $B^{(2)}$ and $A_3^{(2)}$. 
To eliminate this mixing we make the final transformation 
\begin{align}
V^{(7)} = K_7 V^{(8)}, \quad
K_7=
\begin{pmatrix}
1&0&0&0\\
0&1&0&0\\
0&0&1&-s_{\delta_7}\\
0&0&0&c_{\delta_7}
\end{pmatrix},
\end{align}
where $V^{(8)^T}= (D^{(5)}, C^{(5)}, B^{(3)}, A_3^{(3)})$. 
In the basis $V^{(8)}$ the 
kinetic energy for all the gauge fields is in the canonical form  so that
\begin{equation}
\mathcal{L}_{\rm KE}=-\frac{1}{4}(D^{(5)^2}+C^{(5)^2}+ B^{(3)^2}+A_3^{(3)^2}).
\end{equation}
The free Lagrangian in the $V^{(8)}$ basis is then
\begin{align}
\mathcal{L}&= -\frac{1}{4} V^{(8)^T} V^{(8)} - \frac{1}{2} V^{(8)^T} K^T M_D^2 K V^{(8)},
\end{align}
with
\begin{equation}
    K\equiv K_1K_2 K_3 K_4 K_5 K_6 K_7.
\end{equation}

As discussed in the beginning of this section we now make the expansion of Eq.~(\ref{deltams}).
Below we   exhibit $K$ and $\Delta M^2$ in the limit $\delta_1, \delta_2 \ll 1$. In this case
 $s_{\delta_1} \sim \delta_1, c_{\delta_1}\sim 1$, etc.  and $K$ and $\Delta M^2$ have the following 
 form
 \begin{equation}
\label{k8}
K\sim \begin{pmatrix}
1&-\Bar{\delta_2}&-\delta_3&-\delta_4\\
0&1&-\delta_5&-\delta_6\\
0&0&1&-\delta_7\\
0&0&0&1
\end{pmatrix}, ~~
\Delta M^2=
\begin{pmatrix}
0&-m_{\gamma'}^2\Bar{\delta_2}&-m_{\gamma'}^2\delta_3&-m_{\gamma'}^2\delta_4\\
-m_{\gamma'}^2\Bar{\delta_2}&0&-m_{Z'}^2\delta_5&-m_{\gamma'}^2\delta_6\\
-m_{\gamma'}^2\delta_3&-m_{Z'}^2\delta_5&0&0\\
-m_{\gamma'}^2\delta_4&-m_{\gamma'}^2\delta_6&0&0
\end{pmatrix}.
\end{equation}
The interactions relevant for our computation arise from $\Delta M^2$
and  the relation 
\begin{align}
V^{(1)}= K V^{(8)}.
\label{k9}
\end{align}
Eqs.~(\ref{k8}) and~(\ref{k9}) and non-degenerate perturbation theory is utilized in the
computation of couplings of the visible sector with the hidden sector. This  is
discussed in the next section.

\section{Dark photon $\gamma'$ and dark $Z'$ couplings with Standard Model particles}
\label{sec:AppB}

The couplings of the dark photon and $Z'$ to the SM particles are given by the Lagrangian Eq.~(\ref{NC-visible}). 
To compute the couplings proportional to $\delta_1$ and $\delta_2$
 that arise due to the kinetic and Stueckelberg mass
  mixings,  we use first order 
 non-degenerate perturbation theory using $\Delta M^2$  given in  Eq.~(\ref{k8}) as the perturbation. Thus, to first order 
 perturbation in $\Delta M^2$, 
the neutral currents of Eq.~(\ref{NC-visible}) which involve the
 vector and  axial-vector couplings of the dark photon, the dark
 $Z'$ and the SM gauge gauge bosons are given by
\begin{align}
\label{vf}
v_f&=T_{3f}-2Q_f\sin^2\theta_w, \\
\label{af}
a_f&=T_{3f}, \\
\label{vfp}
v'_f&=-Q_f\sin2\theta_w\cos\theta_w(1+\epsilon_z^2)\delta_1\left[1-\left(1-\frac{T_{3f}}{2Q}\right)\frac{m_{Z'}^2}{m^2_W}\right], \\
\label{afp}
a'_f&=-\delta_1 T_{3f} \sin\theta_w \epsilon^2_z(1+\epsilon^2_z), \\
\label{vfpp}
v''_f&=Q_f\sin2\theta_w\cos\theta_w(1+\epsilon_{\gamma'}^2)\left[1-\left(1-\frac{T_{3f}}{2Q}\right)\frac{m_{\gamma'}^2}{m^2_W}\right]\delta_1(\delta_2-\sin\beta), \\
\label{afpp}
a''_f&=T_{3f}\sin\theta_w \epsilon_{\gamma'}^2(1+\epsilon_{\gamma'}^2)\delta_1(\delta_2-\sin\beta),
\end{align}
where $\epsilon_z=m_{Z'}/m_Z$  and $\epsilon_{\gamma'}=m_{\gamma'}/m_Z$.
The relevant couplings with the visible sector are summarized   in Tables~\ref{tab1} and~\ref{tab2}.
Thus Table~\ref{tab1} gives the couplings of $Z$ and $Z'$ to 
the visible sector fermions $f\bar f$ and Table ~\ref{tab2}.
gives the coupling of $\gamma'$ to $f\bar f$.

\begin{table}[tbp]
\centering
\begingroup
\setlength{\tabcolsep}{12pt} 
\renewcommand{\arraystretch}{2.0}
\resizebox{\textwidth}{!}{\begin{tabular}{|cccccc|}
\hline
$f$ & $Q_f$ & $v_f$ & $a_f$ & $v'_f$ & $a'_f$ \\
\hline
$\nu_e,\nu_{\mu},\nu_{\tau}$ & 0 & $\frac{1}{2}$ & $\frac{1}{2}$ & $-\frac{1}{2}\sin\theta_w \epsilon^2_z(1+\epsilon^2_z)\delta_1$ & $-\frac{1}{2}\sin\theta_w \epsilon^2_z(1+\epsilon^2_z)\delta_1$ \\
$e,\mu,\tau$ & $-1$ & $-\frac{1}{2}+2\sin^2\theta_w$ & $-\frac{1}{2}$ & $\sin2\theta_w \cos\theta_w \left(1-\frac{3 m^2_{Z'}}{4 m^2_W}\right)(1+\epsilon^2_z)\delta_1$ & $\frac{1}{2}\sin\theta_w \epsilon^2_z(1+\epsilon^2_z)\delta_1$ \\
$u,c,t$ & $\frac{2}{3}$ & $\frac{1}{2}-\frac{4}{3}\sin^2\theta_w$ & $\frac{1}{2}$ & $-\frac{2}{3}\sin2\theta_w \cos\theta_w \left(1-\frac{5 m^2_{Z'}}{8 m^2_W}\right)(1+\epsilon^2_z)\delta_1$ & $-\frac{1}{2}\sin\theta_w \epsilon^2_z(1+\epsilon^2_z)\delta_1$  \\
$d,s,b$ & $-\frac{1}{3}$ & $-\frac{1}{2}+\frac{2}{3}\sin^2\theta_w$ & $-\frac{1}{2}$ & $\frac{1}{3}\sin2\theta_w \cos\theta_w \left(1-\frac{m^2_{Z'}}{4 m^2_W}\right)(1+\epsilon^2_z)\delta_1$ & $\frac{1}{2}\sin\theta_w \epsilon^2_z(1+\epsilon^2_z)\delta_1$  \\
\hline
\end{tabular}}
\endgroup
\caption{\label{tab1} The $Z\to f\bar{f}$ and $Z'\to f\bar{f}$ vertices. In the above, $\epsilon_z=m_{Z'}/m_Z$ and $m_W$ is the $W$ boson mass.}
\end{table}

\begin{table}[tbp]
\centering
\begingroup
\setlength{\tabcolsep}{12pt} 
\renewcommand{\arraystretch}{2.0}
\resizebox{\textwidth}{!}{\begin{tabular}{|cccc|}
\hline
$f$ & $Q_f$ & $v''_f$ & $a''_f$  \\
\hline
$\nu_e,\nu_{\mu},\nu_{\tau}$ & 0 &  $\frac{1}{2}\sin\theta_w \epsilon^2_{\gamma'}(1+\epsilon^2_{\gamma'})\delta_1(\delta_2-s_{\beta})$ & $\frac{1}{2}\sin\theta_w \epsilon^2_{\gamma'}(1+\epsilon^2_{\gamma'})\delta_1(\delta_2-s_{\beta})$ \\
$e,\mu,\tau$ & $-1$ & $\sin2\theta_w \cos\theta_w \left(1-\frac{3 m^2_{\gamma'}}{4 m^2_W}\right)(1+\epsilon^2_{\gamma'})\delta_1(s_{\beta}-\delta_2)$  &  $-\frac{1}{2}\sin\theta_w \epsilon^2_{\gamma'}(1+\epsilon^2_{\gamma'})\delta_1(\delta_2-s_{\beta})$  \\
$u,c,t$ & $\frac{2}{3}$ &  $\frac{2}{3}\sin2\theta_w \cos\theta_w \left(1-\frac{5 m^2_{\gamma'}}{8 m^2_W}\right)(1+\epsilon^2_{\gamma'})\delta_1(\delta_2-s_{\beta})$  &  $\frac{1}{2}\sin\theta_w \epsilon^2_{\gamma'}(1+\epsilon^2_{\gamma'})\delta_1(\delta_2-s_{\beta})$  \\
$d,s,b$ & $-\frac{1}{3}$ &  $\frac{1}{3}\sin2\theta_w \cos\theta_w \left(1-\frac{m^2_{\gamma'}}{4 m^2_W}\right)(1+\epsilon^2_{\gamma'})\delta_1(s_{\beta}-\delta_2)$  &  $-\frac{1}{2}\sin\theta_w \epsilon^2_{\gamma'}(1+\epsilon^2_{\gamma'})\delta_1(\delta_2-s_{\beta})$   \\
\hline
\end{tabular}}
\endgroup
\caption{\label{tab2} The $\gamma'\to f\bar{f}$ vertices. In the above, $\epsilon_{\gamma'}=m_{\gamma'}/m_Z$.}
\end{table}
The triple gauge boson couplings of $\gamma', Z', Z$ are given by
 are
\begin{align}
&WW\gamma^{\prime}:~-ig_2\cos\theta_w\sin\theta_w(1+\epsilon^2_{\gamma'})\delta_1(\delta_2-\sin\beta).\\
&WWZ^{\prime}:~ig_2\cos\theta_w\sin\theta_w(1+\epsilon^2_z)\delta_1, \\
&WWZ:~-ig_2\cos\theta_w.
\end{align}

\noindent
{\bf Couplings in the limit of large $\delta_2$: }
 Some of the processes such as the lifetime of the dark photon
 require  only that the product $\delta_1\delta_2$ be small
 which could be achieved by $\delta_1$ being small while $\delta_2$
 is $\mathcal{O}(1)$ size. Thus we list below   the vector and axial-vector couplings in the limit of small $\delta_1$ and $\beta$ while $\delta_2$ is not necessarily small
\begin{align}
v_f&=T_{3f}-2Q_f\sin^2\theta_w+\frac{Q_f\delta_1\delta_2\sin\theta_w\sin2\theta_w (\delta_2-\sin\beta+\delta_2^2\sin\beta)}{\sqrt{1-\delta_2^2}}, \\
a_f&=T_{3f}, \\
v'_f&=-Q_f\sin2\theta_w\cos\theta_w(1+\epsilon_z^2+\epsilon^2_{\gamma'}\delta_2^2)\delta_1\left[\frac{m^2_{Z'}}{m^2_{Z'}+m^2_{\gamma'}\delta_2^2}-\left(1-\frac{T_{3f}}{2Q_f}\right)\frac{m_{Z'}^2}{m^2_W}\right] \nonumber \\
&\hspace{4cm}\times\left[(1+\delta_2\sin\beta)+\frac{m^2_{\gamma'}}{m^2_{Z'}}\frac{\delta_2^2(1+\delta_2\sin\beta)-\delta_2\sin\beta}{\sqrt{1-\delta_2^2}}\right], \\
a'_f&=-\delta_1 T_{3f} \sin\theta_w (1+\epsilon^2_z+\delta_2^2\epsilon^2_{\gamma'})\left[\epsilon^2_z(1+\delta_2\sin\beta)+\epsilon^2_{\gamma'}\frac{\delta_2^2(1+\delta_2\sin\beta)-\delta_2\sin\beta}{\sqrt{1-\delta_2^2}}\right], \\
v''_f&=Q_f\sin2\theta_w\cos\theta_w(1+\epsilon_{\gamma'}^2)\left[1-\left(1-\frac{T_{3f}}{2Q_f}\right)\frac{m_{\gamma'}^2}{m^2_W}\right]\frac{\delta_1(\delta_2-\sin\beta+\delta_2^2\sin\beta)}{\sqrt{1-\delta_2^2}}, \\
a''_f&=T_{3f}\sin\theta_w \epsilon_{\gamma'}^2(1+\epsilon_{\gamma'}^2)\frac{\delta_1(\delta_2-\sin\beta+\delta_2^2\sin\beta)}{\sqrt{1-\delta_2^2}}.
\end{align}

The couplings with the $D$ fermions become
\begin{align}
&\gamma' D\bar{D}:~g_X\dfrac{m^2_{\gamma'}\delta_2}{m^2_{Z'}-m^2_{\gamma'}(1-\delta_2^2)}, \\
&Z' D\bar{D}:~g_X, \\
&\gamma D\bar{D}:~-g_X \dfrac{m^2_{\gamma'}\delta_1\delta_2\cos\theta_w}{m^2_{Z'}+m^2_{\gamma'}\delta_2^2}\left[\delta_2(1+\beta\delta_2)+\dfrac{\beta-\delta_2-\beta\delta_2^2}{\sqrt{1-\delta_2^2}}\right], \\
&Z D\bar{D}:~g_X\delta_1\sin\theta_w (1+\epsilon_z^2+\delta_2^2\epsilon^2_{\gamma'})\left[(1+\beta\delta_2)(1+\delta_2^2\epsilon^2_{\gamma'})+\dfrac{\delta_2\epsilon^2_{\gamma'}}{\sqrt{1-\delta_2^2}}(\beta-\delta_2-\beta\delta_2^2)\right],   
\end{align}
and the triple gauge boson couplings take the form
\begin{align}
&WW\gamma^{\prime}:~-ig_2\cos\theta_w\sin\theta_w(1+\epsilon^2_{\gamma'})\frac{\delta_1(\delta_2-s_{\beta}+\beta\delta_2^2)}{\sqrt{1-\delta_2^2}}, \\
&WWZ^{\prime}:~ig_2\cos\theta_w\sin\theta_w\frac{\delta_1(1+\epsilon^2_{z}+\epsilon^2_{\gamma'}\delta_2^2)}{m^2_{Z'}+m^2_{\gamma'}\delta_2^2}\left[m^2_{Z'}(1+\delta_2\beta)+\frac{m^2_{\gamma'}\delta_2(\beta\delta_2^2+\delta_2-\beta)}{\sqrt{1-\delta_2^2}}\right], \\
&WWZ:~-ig_2\cos\theta_w\left[1+\frac{\sin\theta_w\tan\theta_w\delta_1\delta_2(\beta\delta_2^2+\delta_2-\beta)}{\sqrt{1-\delta_2^2}}\right].
\end{align}

\section{Deduction of three temperature Boltzmann equations}
\label{sec:AppC}

Next we give a deduction of the Boltzmann equations for the case of three heat baths. We will use $T_1$ as
the reference temperature. Let us consider a generic  number density $n_i$. In this case
$n_i R^3$ is conserved during the expansion if there is no injection and we have
$\frac{d(n_iR^3)}{dt} =0$ where $R$ is the scale factor,
while in the presence of injection one has
\begin{align}
\frac{dn_i}{dt}+ 3 H n_i=C_i\,,
\label{ni}
\end{align}
where $C_i$ represent the integrated collision terms.
Next, if $S= sR^3$ is the total entropy, it is conserved which implies that
\begin{align}
\frac{ds}{dt} + 3 H s=0.
\label{s}
\end{align}
Using Eqs.~(\ref{ni}) and~(\ref{s}), and the fact that  $n_i= sY_i$, one finds
\begin{align}
\frac{dY_i}{dt} = \frac{1}{s} C_i\,.
\end{align}
We can convert this equation to one that uses temperature $T_1$ rather than time
which gives
\begin{align}
\frac{dY_i}{dT_1}
&= - \frac{d\rho/dT_1}{4 H \rho}\frac{1}{s} C_i\,,
\end{align}
where $d\rho/dT_1$ is given by
\begin{align}
\frac{d\rho}{dT_1}&=(\eta + T_1 \eta')
\frac{d\rho_v}{dT} +
\frac{d\rho_1}{dT_1} +(\zeta+ T_1 \zeta')
 \frac{d\rho_2}{dT_2}.
\end{align}
The Boltzmann equations for $Y_D, Y_{Z'}, Y_{\gamma'}$ may now be written as
\begin{equation}
\begin{aligned}
\frac{dY_D}{dT_1}= -\frac{d\rho/dT_1}{ 4 H \rho}   \frac{1}{s} C_D, \\
\frac{dY_{Z'}}{dT_1}= - \frac{d\rho/dT_1}{ 4 H \rho}   \frac{1}{s} C_{Z'}, \\
\frac{dY_{\gamma'}}{dT_1}= - \frac{d\rho/dT_1}{4 H \rho}   \frac{1}{s} C_{\gamma'},
\end{aligned}    
\end{equation}
where $C_D$, $C_{Z'}$ and $C_{\gamma'}$ are given by
\begin{align}
C_D=&n_i^2(T)\langle\sigma v\rangle_{i\bar i \to D \bar D}(T)+n_{Z'}^2\langle\sigma v\rangle_{Z'Z'\to D\bar D}(T_1)-\frac{1}{2}n_D^2\langle\sigma v\rangle_{D\bar D\to i\bar i}(T_1) \non
&-\frac{1}{2}n_{D}^2\langle\sigma v\rangle_{D\bar D\to Z'Z'}(T_1)-\frac{1}{2}n_D^2\langle\sigma v\rangle_{D\bar D\to \gamma'\gamma'}(T_1)+n_{\gamma'}^2\langle\sigma v\rangle_{ \gamma'\gamma'\to D\bar D}(T_2) \non
&-\frac{1}{2}n_D^2\langle\sigma v\rangle_{D\bar{D}\to Z'\gamma'}(T_1)+n_{Z'}n_{\gamma'}\langle\sigma v\rangle_{Z'\gamma'\to D\bar{D}}(T_1,T_2),
\label{jd}
 \\ \non
C_{Z'}= &n_i^2(T)\langle\sigma v\rangle_{i\bar i \to Z'}(T)+n_i^2(T)\langle\sigma v\rangle_{i\bar i \to Z'Z'}(T)+\frac{1}{2} n_{D}^2\langle\sigma v\rangle_{D\bar D\to Z'Z'}(T_1) \non
&-n_{Z'}^2\langle\sigma v\rangle_{Z'Z'\to i\bar i}(T_1)-n_{Z'}^2\langle\sigma v\rangle_{Z'Z'\to D\bar D}(T_1)-\frac{1}{2}n_D^2\langle\sigma v\rangle_{D\bar{D}\to Z'\gamma'}(T_1) \non
&-n_{Z'}n_{\gamma'}\langle\sigma v\rangle_{Z'\gamma'\to D\bar{D}}(T_1,T_2)-n_{Z'}\langle\Gamma_{Z'\to i\bar i}\rangle(T_1),
 \label{jzprime}
  \\ \non
C_{\gamma'}= &n_i^2(T)\langle\sigma v\rangle_{i\bar i \to \gamma'}(T) +\frac{1}{2}n_D^2\langle\sigma v\rangle_{D\bar D\to \gamma'\gamma'}(T_1)-
n_{\gamma'}\langle\Gamma_{\gamma' \to i\bar i}\rangle(T_2)- n_{\gamma'}^2\langle\sigma v\rangle_{\gamma'\gamma' \to D\bar D}(T_2) \non 
&+n_i^2(T)\langle\sigma v\rangle_{i\bar{i}\to\gamma'\gamma'}(T)-n_{\gamma'}^2\langle\sigma v\rangle_{\gamma'\gamma'\to i\bar{i}}(T_2)+\frac{1}{2}n_D^2\langle\sigma v\rangle_{D\bar{D}\to Z'\gamma'}(T_1) \non
&-n_{Z'}n_{\gamma'}\langle\sigma v\rangle_{Z'\gamma'\to D\bar{D}}(T_1,T_2).
\label{jgammaprime}
\end{align}

In Eqs.~(\ref{jd}),~(\ref{jzprime}) and~(\ref{jgammaprime}) 
one encounters thermally averaged decay width and 
thermally averaged cross sections.
The thermally averaged decay width is given by
\begin{equation}
\langle\Gamma_{a\to bc}\rangle=\Gamma_{a\to bc}\frac{K_1(m_a/T)}{K_2(m_a/T)},
\label{widthav}
\end{equation}
and the thermally averaged cross-section is given by
\begin{equation}
\langle\sigma v\rangle^{a\bar{a}\to bc}(T)=\frac{1}{8 m^4_a T K^2_2(m_a/T)}\int_{4m_a^2}^{\infty} ds ~\sigma(s) \sqrt{s}\, (s-4m_a^2)K_1(\sqrt{s}/T).
\label{sigmav}
\end{equation}
$K_1$ and $K_2$ are the modified Bessel functions of the second kind and degrees one and two, respectively. For the  case when the annihilating particles have different masses $m_1$ and $m_2$ and are at different temperatures $T_1$ and $T_2$, the thermally averaged cross-section becomes
\begin{align}
\langle\sigma v\rangle_{12\to 34}(T_1,T_2)=&\frac{1}{4m_1^2 m_2^2 K_2(m_1/T_1)K_2(m_2/T_2)}\int_{(m_1+m_2)^2}^{\infty}ds\,\sigma(s)\sqrt{s(s-(m_1+m_2)^2)}I(s), 
\label{sigv1234}
\end{align}
where
\begin{equation}
I(s)=\frac{1}{T_1-T_2}\int_{\sqrt{s}}^{\infty}dx\, e^{-a_{+}x/2}\sinh\left(\frac{a_{-}}{2}\sqrt{1-\frac{(m_1+m_2)^2}{s}}\sqrt{x^2-s}\right),     
\end{equation}
and where
\begin{equation}
a_{+}=\frac{T_1+T_2}{T_1 T_2},~~~ a_{-}=\frac{T_1-T_2}{T_1 T_2}.   
\end{equation}
Note that in the limit $T_1\to T_2$, $I(s)\to \dfrac{\sqrt{s-(m_1+m_2)^2}}{2T}K_1(\sqrt{s}/T)$ which for $m_1=m_2$ allows us to recover Eq.~(\ref{sigmav}) using Eq.~(\ref{sigv1234}).
The equilibrium yield of species $i$ is given by
\begin{equation}
Y_i=\frac{n_i^{\rm eq}}{s}=\frac{g_i}{2\pi^2 s}m_i^2 T K_2(m_i/T).
\end{equation}
The hidden sectors degrees of freedom is given by
\begin{align}
\label{g1eff}
g_{1\rm eff}&= g^{Z'}_{\rm eff} +\frac{7}{8}  g^D_{\rm eff},~~~\text{and}~~~h_{1\rm eff}= h^{Z'}_{\rm eff} + \frac{7}{8} h^D_{\rm eff}, \\
g_{2\rm eff}&= g^{\gamma'}_{\rm eff},~~~\text{and}~~~h_{2\rm eff}= h^{\gamma'}_{\rm eff}, 
\label{g2eff}
\end{align}
where
\begin{equation}
\begin{aligned}
g^{V}_{\rm eff}& = \frac{45}{\pi^4} \int_{x_{V}}^{\infty} \frac{\sqrt{x^2-x_{V}^2} }{e^x-1 } x^2 dx,~~~\text{and}~~~h^{V}_{\rm eff}= \frac{45}{4\pi^4} \int_{x_{V}}^{\infty} \frac{\sqrt{x^2-x_{V}^2} }{e^x-1 } 
(4x^2-x_{V}^2) dx, \\
g^{D}_{\rm eff}& = \frac{60}{\pi^4} \int_{x_{D}}^{\infty} \frac{\sqrt{x^2-x_{D}^2} }{e^x+1 } x^2 dx,~~~\text{and}~~~h^{D}_{\rm eff}= \frac{15}{\pi^4} \int_{x_{D}}^{\infty} \frac{\sqrt{x^2-x_{D}^2} }{e^x+1 } 
(4x^2-x_D^2) dx.
\end{aligned}
\label{hdof}
\end{equation}
In Eq.~(\ref{hdof}), $V=Z',\gamma'$ and  we take $g_{\gamma'}=g_{Z'}=3$ and $g_D=4$.

\section{Dark photon and dark fermion scattering cross sections and $Z'$ decay width}
\label{sec:AppD}

The calculation of the relic densities of the dark photon and dark fermion require solving the coupled Boltzmann equations which contain a variety of cross sections involving the
standard model and dark sector particles. We list these below.

\begin{enumerate}

\item Processes: $D\bar{D}\to Z,Z',\gamma'\to f\bar{f}$

\begin{align}
\sigma^{D\bar{D}\to f\bar{f}}(s)&=\frac{g_X^2 g_2^2 N_c}{12\pi\cos^2\theta_w}\left(1+\frac{2m_D^2}{s}\right)\sqrt{\frac{s-4m_f^2}{s-4m_D^2}}\Bigg\{\frac{[a_f''^2(s-4m^2_f)+v_f''^2(s+2m^2_f)]\delta_2^2}{\kappa^2[(s-m^2_{\gamma'})^2+m^2_{\gamma'}\Gamma^2_{\gamma'}]} \nonumber \\
&+\frac{a_f'^2(s-4m^2_f)+v_f'^2(s+2m^2_f)}{(s-m^2_{Z'})^2+m^2_{Z'}\Gamma^2_{Z'}}+\frac{[a_f^2(s-4m^2_f)+v_f^2(s+2m^2_f)]\delta_1^2\sin^2\theta_w}{(s-m^2_{Z})^2+m^2_{Z}\Gamma^2_{Z}} \nonumber \\
&+\frac{2\delta_2[a_f' a_f''(s-4m^2_f)+v_f' v_f''(s+2m^2_f)]}{\kappa[(s-m_{Z'}^2)^2+m^2_{Z'}\Gamma^2_{Z'}][(s-m_{\gamma'}^2)^2+m^2_{\gamma'}\Gamma^2_{\gamma'}]} G(s,m_{Z'},m_{\gamma'})\nonumber \\
&+\frac{2\delta_1\delta_2[a_f a_f''(s-4m^2_f)+v_f v_f''(s+2m^2_f)]\sin\theta_w}{\kappa[(s-m_{Z}^2)^2+m^2_{Z}\Gamma^2_{Z}][(s-m_{\gamma'}^2)^2+m^2_{\gamma'}\Gamma^2_{\gamma'}]}G(s,m_{Z},m_{\gamma'}) \nonumber \\
&+\frac{2\delta_1[a_f a_f'(s-4m^2_f)+v_f v_f'(s+2m^2_f)]\sin\theta_w}{[(s-m_{Z'}^2)^2+m^2_{Z'}\Gamma^2_{Z'}][(s-m_{Z}^2)^2+m^2_{Z}\Gamma^2_{Z}]}G(s,m_{Z},m_{Z'})\Bigg\},
\end{align}
Here $s$ is the Mandelstam variable which gives the square of the total energy in the CM system. Further, the notation used above is as follows:
 $f=e,\mu,\tau$: $T_{3f}=-1/2$ and $Q_f=-1$ and for $f=u,c,t$: $T_f^3=1/2$ and $Q_f=2/3$ and for $f=d,s,b$: $T_f^3=-1/2$ and $Q_f=-1/3$, $N_c$ is the color number and  
\begin{align}
\kappa&=-(1-m^2_{Z'}/m^2_{\gamma'}), \\
G(s,m_1,m_2)&=(s-m^2_1)(s-m^2_2)+\Gamma_1\Gamma_2 m_{1}m_{2}. 
\end{align}

\item Processes: $D\bar{D}\to Z,Z',\gamma'\to\nu\bar{\nu}$
\begin{align}
\sigma^{D\bar{D}\to \nu\bar{\nu}}(s)=&\frac{g_X^2g_2^2\delta_1^2}{8\pi}\frac{(s+4m_D^2)\tan^2\theta_w}{(1-4m^2_D/s)^{1/2}}\left[\frac{A}{(s-m^2_{Z'})(s-m^2_{\gamma'})}-\frac{1}{(s-m^2_Z)}\right]^2,
\end{align}
where
\begin{equation}
A=\epsilon^2_z(s-m^2_{\gamma'})+\frac{\delta_2\epsilon^2_{\gamma'}(s-m^2_{Z'})(\delta_2-\sin\beta)}{1-m^2_{Z'}/m^2_{\gamma'}}.
\end{equation}

\item Processes: $f\bar{f}\to Z,Z',\gamma'\to D\bar{D}$
\begin{equation}
(s-4m^2_D)\sigma^{D\bar{D}\to f\bar{f}}(s)=N_c^2(s-4m^2_f)\sigma^{f\bar{f}\to D\bar{D}}(s).
\end{equation}

\item Processes: $\nu\bar{\nu}\to Z,Z',\gamma'\to D\bar{D}$
\begin{equation}
\sigma^{\nu\bar{\nu}\to D\bar{D}}(s)=4\left(1-\frac{4m^2_D}{s}\right)\sigma^{D\bar{D}\to \nu\bar{\nu}}(s).
\end{equation}

\item Process: $D\bar{D}\to Z'Z' $ 
\begin{align}
\sigma^{D\bar{D}\to Z'Z'}(s)&=\frac{g_X^4}{8\pi s}\Bigg\{-\sqrt{\frac{s-4m^2_{Z'}}{s-4m^2_D}}\left[\frac{m^2_D s+2m^4_{Z'}+4m^4_D}{(s-4m^2_{Z'})m^2_D+m^4_{Z'}}\right] \nonumber \\
&+\frac{s^2+4m^2_D(s-2m^2_{Z'})+4m^4_{Z'}-8m^4_D}{(s-2m^2_{Z'})(s-4m^2_D)}\log B\Bigg\},
\end{align}
where
\begin{equation}
B=\frac{s-2m^2_{Z'}+\sqrt{(s-4m^2_{Z'})(s-4m^2_D)}}{s-2m^2_{Z'}-\sqrt{(s-4m^2_{Z'})(s-4m^2_D)}}.
\end{equation}

\item Process: $D\bar{D}\to\gamma'\gamma'$ 
\begin{equation}
\sigma^{D\bar{D}\to\gamma'\gamma'}(s)=\left(\frac{\delta_2 M^2_{\gamma'}}{M^2_{Z'}-M^2_{\gamma'}}\right)^4\sigma^{D\bar{D}\to Z'Z'}(s)\Big|_{m_{Z'}\longleftrightarrow m_{\gamma'}}.
\end{equation}

\item Processes: $D\bar{D}\to VV$ with $V=Z',\gamma'$
$$8(s-4m^2_D)\sigma^{D\bar{D}\to VV}(s)=9(s-4m^2_{V})\sigma^{VV\to D\bar{D}}(s).$$

\item Processes $D\bar{D}\rightarrow Z'\gamma'$
\begin{align}
\sigma^{D\bar{D}\rightarrow Z'\gamma'}(s)&= \frac{\delta_2^2g_X^4 m_{\gamma'}^4}{4\pi(m_{Z'}^2-m_{\gamma'}^2) s(s-4m_D^2)} \non
&\times\bigg\{\frac{4m_D^4s-2m_{Z'}m_{\gamma'}s+m_D^2\left[(m_{Z'}^2-m_{\gamma'}^2)^2+s^2\right]}{m_{Z'}^2m_{\gamma'}^2s+m_{D}^2\left[m_{\gamma'}^4+(s-m_{Z'}^2)^2-2m_{\gamma'}^2(s+m_{Z'}^2)\right]}E \non
&+\left[-8m_D^4+(m_{Z'}^2+m_{\gamma'}^2)^2-4m_D^2(m_{Z'}^2-m_{\gamma'}^2-s)+s^2\right]\log F\bigg\},
\end{align}
where
\begin{align}
E&=\sqrt{\left(1-\frac{4m_D^2}{s}\right)\left[m_{\gamma'}^4+(s-m_{Z'}^2)^2-2m_{\gamma'}^2(s+m_{Z'}^2)\right]}, \\
F&=\frac{m_{\gamma'}^2+m_{Z'}^2-s+E}{m_{\gamma'}^2+m_{Z'}^2-s-E}.
\end{align}

\item Processes $Z'\gamma'\rightarrow D\bar{D}$
\begin{equation}
    \sigma^{Z'\gamma'\rightarrow D\bar{D}}(s)=\frac{4s(s-4m_D^2)}{9[(m_{\gamma'}^2+m_{Z'}^2-s)^2]-4m_{\gamma'}^2m_{Z'}^2}\sigma^{D\bar{D}\rightarrow Z'\gamma'}(s)
\end{equation}
\item Processes: $VV\to f\bar{f}$ with $V=Z',\gamma'$

\begin{align}
\sigma^{VV\to f\bar{f}}(s)&=\frac{g_2^4 N_c}{9\pi m^4_{V}s(s-4m^2_{V})\cos^4\theta_w}\Bigg\{\frac{\sqrt{(s-4m^2_f)(s-4m^2_{V})}}{m^4_{V}+m^2_f(s-4m^2_{V})}\nonumber \\
&\times \Big(c_A^4 [-2 m_{V}^8 + m_f^2 m_{V}^4 (s+4 m_{V}^2) + 2 m_f^4 (8 m_{V}^4 - 8 m_{V}^2 s + s^2)] \nonumber \\
&~~~~+2 c_A^2 c_V^2 m_{V}^4 [8 m_f^4 - 6 m_{V}^4 + m_f^2 (22 m_{V}^2-7 s)]\nonumber \\
&~~~~-m_{V}^4 (4 m_f^4 + 2 m_{V}^4 + m_f^2 s) c_V^4\Big)\nonumber \\
&+\frac{\log C}{(s-2m^2_{V})}\Big(c_A^4 [4 m_f^4 (4 m_{V}^2 - s) s +m_{V}^4 (4 m_{V}^4 + s^2) \nonumber \\
&~~~~~~~~~~~~~~~~~~~+4 m_f^2 m_{V}^2 (-4 m_{V}^4-3 m_{V}^2 s + s^2)]+2 c_A^2 m_{V}^2 [16 m_f^4 m_{V}^2 \nonumber \\
&~~~~~~~~~~~~~~~~~~~+3 m_{V}^2 (s^2+4 m_{V}^4)+2 m_f^2 (s^2-10 m_{V}^2 s-10 m_{V}^4)] c_V^2 \nonumber \\
&~~~~~~~~~~~~~~~~~~~+m_{V}^4 [s^2 + 4 m_f^2 (s-2 m_{V}^2)-8 m_f^4 + 4 m_{V}^4] c_V^4\Big)\Bigg\},
\end{align}
where
\begin{equation}
C=\frac{s-2m^2_{V}+\sqrt{(s-4m^2_{V})(s-4m^2_f)}}{s-2m^2_{V}-\sqrt{(s-4m^2_{V})(s-4m^2_f)}},
\end{equation}
and $c_A=a_f'$, $c_V=v_f'$ for $V=Z'$ and $c_A=a_f''$, $c_V=v_f''$ for $V=\gamma'$.  \\

\item Process: $VV\to\nu\bar{\nu}$ with $V=Z',\gamma'$
\begin{align}
\sigma^{VV\to\nu\bar{\nu}}(s)&=\frac{g_2^4 \delta_1^4}{6\pi s}\tan^4\theta_w\Bigg[\frac{s^2+4m^4_{V}}{(s-4m^2_{V})(s-2m^2_{V})}\log D -2\left(1-\frac{4m^2_{V}}{s}\right)^{-1/2}\Bigg]\nonumber \\
&\hspace{3cm}\times
\begin{cases}
\epsilon_z^8,& \text{for } V=Z' \\
 (\delta_2-\sin\beta)^4\epsilon_{\gamma'}^8, &\text{for } V=\gamma' 
\end{cases},
\end{align}
where
\begin{equation}
D=\frac{s-2m^2_{V}+\sqrt{s(s-4m^2_{V})}}{s-2m^2_{V}-\sqrt{s(s-4m^2_{V})}}.
\end{equation}

\item Processes $f\bar{f},\nu,\bar{\nu}\to VV$ with $V=Z',\gamma'$
\begin{align}
9(s-4m^2_V)\sigma^{VV\to f\bar{f}}&=8(s-4m^2_f)\sigma^{f\bar{f}\to VV}, \\
9(s-4m^2_V)\sigma^{VV\to \nu\bar{\nu}}&=2s\,\sigma^{\nu\bar{\nu}\to VV}.
\end{align}

\item Process: $f\bar{f}\to Z'$
\begin{align}
\sigma^{f\bar{f}\to Z'}(s)&=\frac{\pi g_2^2\delta_1^2 m_{Z'}^2}{2s\sqrt{s-4m^2_f} N_c}\Bigg[\left(1+\frac{2m^2_f}{m_{Z'}^2}\right)\left(Q_f+\frac{m_{Z'}^2}{2m_W^2}(T_{3f}-2Q_f)\right)^2\sin^2 2\theta_w \nonumber \\
&\hspace{3.5cm}+\left(1-\frac{4m^2_f}{m^2_{Z'}}\right)\epsilon_z^4 T^2_{3f}\tan^2\theta_w\Bigg]\delta(\sqrt{s}-m_{Z'}).
\end{align}

\item Process: $\nu\bar{\nu}\to Z'$
\begin{align}
\sigma^{\nu\bar{\nu}\to Z'}(s)&=\frac{3\pi g_2^2\delta_1^2\epsilon_z^4 m^2_{Z'}}{s^{3/2}}\tan^2\theta_w\,\delta(\sqrt{s}-m_{Z'}).
\end{align}

\item Process: $f\bar{f},\nu\bar{\nu}\to \gamma'$
\begin{align}
\sigma^{f\bar{f}\to\gamma'}(s)=(\delta_2-\sin\beta)^2\,\sigma^{f\bar{f}\to Z'}(s)\Big|_{m_{Z'}\longleftrightarrow m_{\gamma'}}.
\end{align}
Same applies to $\sigma^{\nu\bar{\nu}\to\gamma'}(s)$.

\item Process: $Z'\to f\bar{f},\nu\bar{\nu}$

The decay width of $Z'$ to SM fermions is given by
\begin{align}
\Gamma_{Z'\to f\bar{f}}&=\frac{g_2^2\delta_1^2 N_c}{12\pi}m_{Z'}\sqrt{1-\left(\frac{2m_f}{m_{Z'}}\right)^2}\Bigg[\left(1-\frac{4m^2_f}{m^2_{Z'}}\right)T^2_{3f}\epsilon^4_z \tan^2\theta_w \nonumber \\
&+\left(1+\frac{2m^2_f}{m^2_{Z'}}\right)\left(Q_f+\frac{m_{Z'}^2}{2m_W^2}(T_{3f}-2Q_f)\right)^2\sin^2 2\theta_w\Bigg],
\end{align}
and its invisible decay is
\begin{align}
\Gamma_{Z'\to \nu\bar{\nu}}&=\frac{g_2^2\delta_1^2\epsilon_z^4}{8\pi}m_{Z'}\tan^2\theta_w.
\end{align}

\end{enumerate}

\end{document}